\documentclass{aa}
\usepackage{graphicx}
\usepackage{booktabs}
\usepackage{epsfig}
\usepackage{txfonts}
\newcommand{\mincir}{\raise
-3.truept\hbox{\rlap{\hbox{$\sim$}}\raise4.truept\hbox{$<$}\ }}
\begin{document}

   \title{XXL Survey XXI}
     \subtitle{The environment and clustering of X-ray AGN in the XXL-South field}

 % \subtitle{O.Melnyk et al. The XXL Survey. XXI.}

   \author{O. Melnyk\inst{1,2}\and A. Elyiv\inst{2}\and V. Smol\v{c}i\'{c}\inst{1}\and  M. Plionis\inst{3,4,5}\and E. Koulouridis\inst{6}\and S. Fotopoulou\inst{7}\and
      L. Chiappetti\inst{8}\and C. Adami\inst{9}\and N. Baran\inst{1}\and A. Butler\inst{10}\and J. Delhaize\inst{1}\and I. Delvecchio\inst{1}\and F. Finet\inst{11}\and M. Huynh\inst{10}\and 
      C. Lidman\inst{12}\and M. Pierre\inst{6}\and E. Pompei\inst{13}\and  C. Vignali\inst{14,15} \and J. Surdej\inst{16}}

   \offprints{O. Melnyk}

  \institute{Department of Physics, Faculty of Science, University of Zagreb, Bijeni\v{c}ka cesta 32, 1000 Zagreb, Croatia\\
  \email{melnykol@gmail.com}
\and
Main Astronomical Observatory, Academy of Sciences of Ukraine, 27 Akademika Zabolotnoho St., 03680 Kyiv, Ukraine \and
    Physics Dept., Aristotle Univ. of Thessaloniki, Thessaloniki 54124, Greece \and
  Instituto Nacional de Astr\'ofisica, \'Optica y Electr\'onica, 72000 Puebla, Mexico \and
IAASARS, National Observatory of Athens, GR-15236 Penteli, Greece \and
DSM/Irfu/SAp, CEA/Saclay, F-91191 Gif-sur-Yvette Cedex, France \and
Department of Astronomy, University of Geneva, ch. d'Ecogia 16, 1290 Versoix, Switzerland \and
INAF-IASF Milano, via Bassini 15, I-20133 Milano, Italy \and
Aix Marseille Univ., CNRS, LAM, Laboratoire d'Astrophysique de Marseille, Marseille, UMR 7326, 13388, Marseille, France \and
International Centre for Radio Astronomy Research, M468, University of Western Australia, Crawley, WA 6009, Australia \and
Subaru Telescope, National Astronomical Observatory of Japan, 650 North Aohoku Place, Hilo, HI 96720 \and
Australian Astronomical Observatory, North Ryde, NSW 2113, Australia \and
European Southern Observatory, Alonso de Cordova 3107, Vitacura, 19001 Casilla, Santiago 19, Chile \and
Dipartimento di Fisica e Astronomia, Alma Mater Studiorum, Universit\`a degli Studi di Bologna, Via Gobetti 93/2, 40129 Bologna, Italy \and
INAF -- Osservatorio Astronomico di Bologna, Via Gobetti 93/3, 40129 Bologna, Italy \and
Institut d'Astrophysique et de G\'eophysique, Universit\'e de Li\`ege,  All\'ee du 6 Ao\^ut 19c, 4000 Li\`ege, Belgium  \\
}
%} 

%              \thanks{}

\date{Received Xxxxx xx, 2010; accepted Xxxx xx, 2010}

% \abstract{}{}{}{}{}
% 5 {} token are mandatory

 \abstract
   {This work is part of a series of studies focusing on the environment and the properties of the X-ray selected active galactic nuclei (AGN) population from the XXL survey. 
   The present survey, given its large area,    continuity, extensive multiwavelength coverage, and large-scale structure  information, is ideal for this kind of study. 
   Here, we focus on the XXL-South  (XXL-S) field.     } 
   %aims heading (mandatory)
   {Our main aim is to study the environment of the various types of X-ray selected AGN and investigate its possible role in AGN triggering and evolution.}
 %  methods heading (mandatory)
  {We studied the large-scale ($>1$ Mpc) environment up to redshift $z=1$ using the nearest neighbour distance method to compare various pairs of AGN types.  
  We also investigated the small-scale environment ( $<0.4$ Mpc) by calculating the local overdensities of optical galaxies. 
  In addition, we built a catalogue  of AGN concentrations with two or more members using the hierarchical clustering method and we correlated them with 
  the X-ray galaxy clusters detected in the XXL survey.}
  {It is found that radio detected X-ray sources are more obscured than non-radio ones, though not all  radio sources are obscured AGN. 
  We did not find any significant differences in the large-scale clustering between luminous and faint X-ray  AGN, or between obscured and unobscured ones, 
  or between radio and non-radio sources. At local scales ($<0.4$ Mpc), AGN typically reside in overdense regions, compared to non-AGN; 
  however, no differences were found between the various types of AGN. A majority of AGN concentrations with two or more members are found in the neighbourhood 
  of X-ray galaxy clusters within $<$25-45 Mpc. Our results suggest that X-ray AGN are typically located in supercluster filaments, 
  but they are also found in over- and underdense regions.}
{}
     {}


   \maketitle
%
%________________________________________________________________

\section{Introduction}

Over the last decade many authors have been studying the local and large-scale environment of active galactic nuclei (AGN), investigating a possible link between nuclear activity, 
host galaxy properties and environment. 
For example \cite{2009MNRAS.392.1509G},  \cite{2008ApJ...673..715C}, Silverman et al. (2009), Lietzen et al. (2009, 2011),  Tasse et al. (2008, 2011), 
\cite{2013A&A...557A..81M}, 
\cite{2013MNRAS.430.3086G}, and Karouzos et al. (2014a,b) have shown that
AGN reside in any type of environment, including underdense
and overdense regions. Gilli et al. (2009) did not find any evidence that AGN with broad optical lines cluster differently from AGN without broad optical lines. 

On the other hand, others have reported environmental differences 
between various types of AGN (obscured, unobscured, luminous, faint, FRI, FRII, etc.). 
In particular, \cite{2006ApJ...639...37K} and \cite{2016A&A...596A..20V} found that 
Seyfert-2 galaxies (Sy2) have  close neighbours more frequently than Seyfert-1 galaxies (Sy1). In addition, these neighbours present evidence of more recent 
interactions than their Sy1 peers (Koulouridis et al. 2013). Melnyk et al. (2013) found evidence that hard X-ray (obscured) 
AGN are located in more overdense regions than soft X-ray (unobscured) ones. Koulouridis et al. (2014) reported a significantly higher frequency 
of merging for non-hidden broad-line region (HBLR) Sy2s, than for HBLR ones. 
In an earlier study, \cite{2008ApJ...688..180S} showed 
that optical luminous AGN inhabit denser environments than  low-luminosity ones (however, see Karouzos et al. 2014a). 

Gandhi et al. (2006) and Gendre et al. (2013) showed  that independently from the radio excitation mode, 
FRI sources are found to lie in higher density environments than FRII sources.
Sabater et al. (2013) outlined the importance of both small-scale and large-scale environmental influence on AGN properties. Low excitation radio galaxies 
(LERG)  or low-to-moderate radiative luminosity AGN (MLAGN) are found in more dense regions than high excitation radio galaxies 
(HERG) or moderate-to-high radiative luminosity AGN (HLAGN);  the modern classification of radio sources can be found in
Smol\v{c}i\'{c} et al. (2017). Sabater et al. (2013) explained
this by the presence of warmer gas at higher densities than is accreted
at low rates in a radiatively inefficient manner, triggering
typical low-luminosity radio AGN. The fraction of
HERG and optical AGN increases with increasing one-on-one interactions, which can funnel cold gas to the
nuclear regions. Similar results were obtained by \cite{2015MNRAS.453.2682I}. 

Differences were also reported for differently selected AGN (optical, IR, radio, or X-ray).
Koutoulidis et al. (2013) concluded that X-ray selected AGN reside in 
significantly more massive dark-matter host halos than optically selected ones. Hickox et al. (2011) and Elyiv et al. (2012) 
found stronger clustering of obscured than unobscured quasar (QSO), 
contrary to \cite{2011ApJ...736...99A} and \cite{2014ApJ...796....4A}. However, their samples were differently selected.
Karouzos et al. (2014a) compared the environmental properties of X-ray, radio, and IR-selected AGN, 
and found that X-ray selected AGN reside in more dense environments; radio AGN also prefer overdense regions, but they can be found in a variety of environments. 
However, a small population of the most luminous radio sources was found in an overdense environment, 
while the most radio-faint ones were found in underdense regions; IR-selected AGN were found in very local overdensities.
Mendez et al. (2016) found that X-ray selected sources are more clustered than IR-selected ones, and linked this difference 
to their distinct host-galaxy populations.
Tasse et al. (2008, 2011) reported that X-ray selected type 2 AGN are located in underdense regions 
similarly to low-mass radio-loud AGN. However,  the high-mass radio-loud AGN prefer overdense regions. Up-to-date observational properties of all AGN types in different 
electromagnetic bands are summarized in Padovani et al. (2017).

The simplest version of the unified scheme by \cite{1993ARA&A..31..473A} proposed that different types of AGN, like Seyfert-1 and Seyfert-2 galaxies, 
as well as broad- and narrow-line AGN/QSOs (i.e. unobscured and obscured AGN) are intrisically the same objects,  the only difference being the 
orientation of an obscuring torus with respect to our line of sight. Therefore, they should present the same environmental properties. 
However, the reason for the differences reported in the above-mentioned studies might be due to one of the following: 1) selection and observational effects, 2) the intrinsically 
different properties of different AGN types, or 3) the different evolutionary stages of different AGN types. If   cases 2 and 3 are true, the simplest version of the unified scheme
cannot fully describe the observational properties of all the different sources and needs to be refined. 

To better understand the above issues, the present work is dedicated to the analysis of the small- and large-scale environment and the clustering properties
of different types of X-ray selected point-like sources (mainly AGN) in the 25 deg$^2$ XXL-South   (XXL-S) field. 
This field is now characterized by a homogeneous optical spectroscopic coverage up to $m_r$ = 21.8 owing to two recently dedicated spectroscopic campaigns 
(Lidman et al. 2016, XXL Paper XIV, and Chiappetti et al., XXL Paper XXVII). 
More than 4500 redshifts are available for the optical counterparts of X-ray point-like sources, which mainly consist of AGN. 
The environmental properties of these AGN are reported for the first time in this field.

In \S2 we present the sample. In \S3, we
describe the nearest neighbour analysis, while in \S4 we present the methodology and the results of the
optical overdensity analysis around X-ray selected point-like sources. The method 
of X-ray galaxy cluster matching with AGN agglomerates is described in \S5. 
Discussion of the main results and a summary are presented in \S6. Throughout
this work we use the standard cosmology: $\Omega_{0}$=0.27, $\Omega_{\Lambda}$=0.73, and
$H_{0}$=71 km/s/Mpc.

\section {The XXL Survey and the sample}
The Ultimate XMM Extragalactic Survey (XXL) is an international project based on the XMM-Newton Very Large Programme surveying $\sim$50 deg$^2$ of the extragalactic 
sky (Pierre et al. 2016, hereafter XXL Paper I). 
The XXL survey contains two nominally equal fields at a depth of $\sim4\times 10^{-15}$ erg cm$^{-2}$ s$^{-1}$ and $\sim2\times 10^{-14}$ erg cm$^{-2}$ s$^{-1}$ 
in the soft (0.5-2.0 keV) and hard (2.0-10.0 keV) bands, 
respectively (\cite{2016A&A...592A...1P}, Figure 4). It comprises 622 XMM pointings with a
total exposure $>$6Ms, and a median exposure of 10.4 ks (see \cite{2016A&A...592A...1P} for details). The two fields have an extensive multiwavelength coverage 
from X-ray to radio wavelengths (the detailed descriptions are given in \cite{2016A&A...592A...1P} and Fotopoulou et al. 2016, hereafter \cite{2016A&A...592A...5F}\footnote
{The multiwavelength and spectroscopic information of the XXL is summarized on the web page http://xxlmultiwave.pbworks.com.}). 
Baran et al. (2016; hereafter \cite{2016A&A...592A...8B}) and Smol{\v c}i{\'c} et al. (2016; hereafter \cite{2016A&A...592A..10S}) presented radio observations of XXL with the 
Very Large Array (VLA) and the Australia Telescope Compact Array (ATCA), respectively. 
Butler et al. 2017 (hereafter \cite{XXL Paper XVIII}) reported the new observations of the full XXL-S  with ATCA.

\begin{figure}
\includegraphics[width=8.5cm,trim=0 0 5 5,clip]{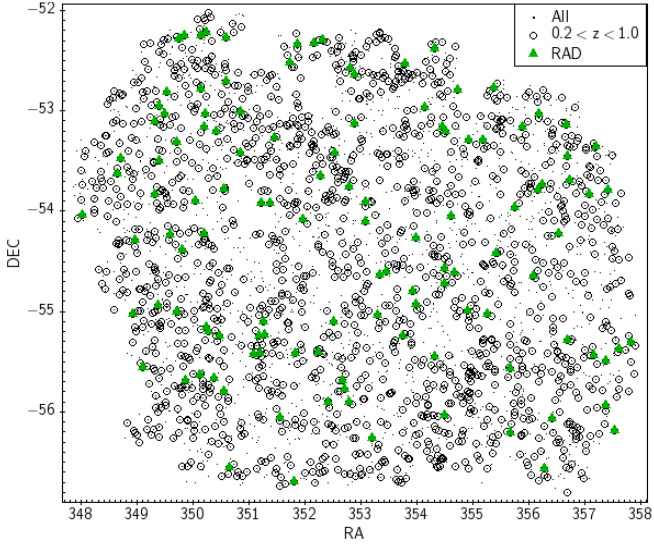}
\caption{Distribution of the XXL-S X-ray point-like sources with optical counterparts considering all the redshifts (N=3280; black dots) 
and sources with spectroscopic redshifts in the range 0.2 $< z <$ 1.0 (N=1592; open circles). 
The positions of the radio-selected X-ray point-like sources (RAD; N=270) are marked with green triangles.}
         \label{1}
      \end{figure}

The XXL-S  field is one of the two XXL fields,  centred at RA=23$^{h}$30 and DEC=-55$^{d}$00. 
It occupies an area of $\sim$25 deg$^2$ containing 11316 X-ray point-like sources\footnote
{We considered here only point-like sources from the ``good'' XMM pointings, i.e. with the condition Xbadfield$<$3, http://cosmosdb.iasf-milano.inaf.it/XXL/}. 
Figure 1 presents the spatial distribution of the X-ray point-like sources with optical counterparts from the Blanco Cosmology Survey\footnote
{http://www.usm.uni-muenchen.de/BCS/} (BCS; Desai et al. 2012). Objects in the 0.2$<z<$1.0 redshift range,
which constitute the target sample of this paper, and the radio sources among them are marked differently on the map.  
The description of the X-ray--multiwavelength associations is presented in \cite{2016A&A...592A...5F}.
The spectroscopic redshifts were taken from \cite{2016PASA...33....1L}), and have been significantly enriched (52\%) 
by new observations taken in 2016, described in XXL Paper XXVII, and  obtained with the AAOmega spectrograph on the Anglo-Australian Telescope. 
A few tens of redshifts were taken from the Marseille CeSaM database\footnote{http://cesam.oamp.fr/xmm-lss/}, 
obtained in the frame of the ESO follow-up programme with FORS2\footnote{FOcal Reducer and low dispersion Spectrograph for the Very Large Telescope (VLT) 
of the European Southern Observatory (ESO)}. We only considered those high-quality spectroscopic redshifts located within 1.1 arcsec around the 
optical counterparts. 
The redshift distribution of the sample of 3280 objects is shown in Figure 2.

\begin{figure}
\includegraphics[width=8.5cm,trim=0 0 5 5,clip]{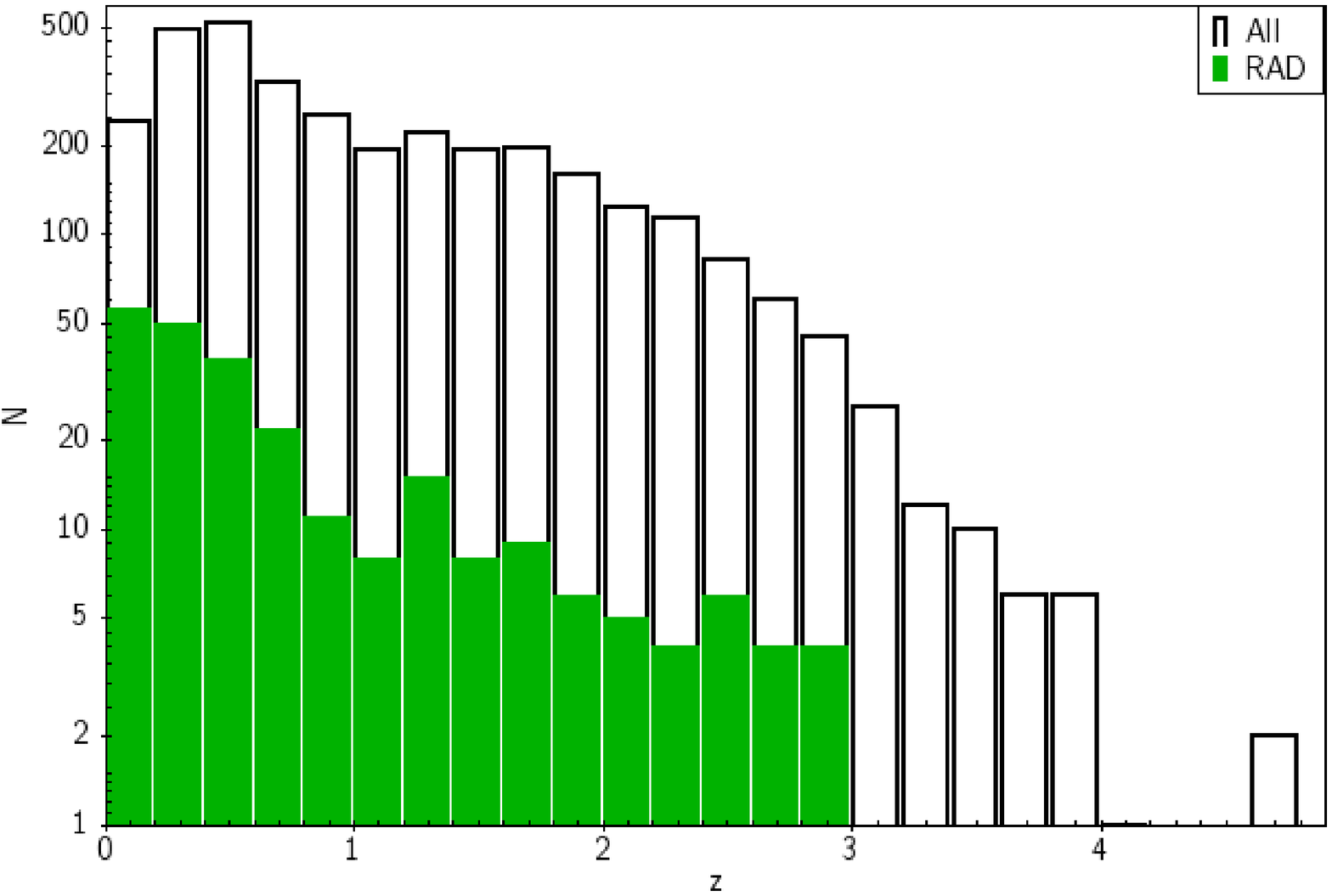}
\caption{Redshift distribution of the XXL-S X-ray point-like sources with spectroscopic (N=3280) redshifts. 
The green histogram represents the redshift distribution of the radio-selected X-ray point-like sources (RAD; N=270).}
\label{2}
\end{figure}

\begin{table}
\caption{Description of the various subsamples of the X-ray point-like source population (see \S2 for details).
The indicated number of sources refers to the samples in the 0.2$<z<$0.6 and 0.6$<z<$1.0 redshift ranges.}
\tabcolsep 2.5 pt
\centering
\begin{tabular}{|l|c|c|c|} \hline
Sample & $N_{0.2<z<0.6}$ & $N_{0.6<z<1.0}$ & Description \\   \hline      
Total &  1012 & 580  & all sources in the entire redshift range   \\ \hline
GAL &  107 & 0  & $L_{X, hard} < 2\times 10^{42}$ erg/s   \\
&  & & if no hard band $L_{X, soft} < 10^{42}$ erg/s \\ \hline
AGN & 905 & 580  & $L_{X, hard} > 2\times 10^{42}$ erg/s  \\ 
& & & if no hard band $L_{X, soft} > 10^{42}$ erg/s  \\ \hline
AGN$_{soft}$    &       626     & 486 & AGN with $HR<$-0.2      \\
AGN$_{hard}$    &       279     & 94 & AGN with $HR\geq$-0.2    \\   \hline
RAD    & 88 & 33  & X-ray sources with radio counterparts   \\ 
 & &  & from \cite{XXL Paper XVIII} catalogue \\ 
nRAD &    924 & 547 & sources without radio counterparts  \\ \hline
Faint   &       145     & 89    & 1/4 of the sample     \\ 
 AGN & & & with lowest $L_{X, hard}$ luminosities \\ \hline
Lum.    &       140     &  89   & 1/4 of the sample     \\ 
AGN & & & with highest $L_{X, hard}$ luminosities \\ 

\hline
\end{tabular}

\end{table}

For the purposes of the present study we divided our sample into various subsamples as described below 
and considered them in two redshift ranges: 0.2$<z<$0.6 and 0.6$<z<$1.0. Table 1 contains a brief description of each sample.

{\bf GAL and AGN}. We separated the full sample into two subsamples, X-ray galaxies (GAL) and AGN. We considered a source as an AGN if $L_{X, hard} > 2\times 10^{42}$ erg/s  
(or $L_{X, soft} > 10^{42}$ erg/s if the X-ray source was detected only in the soft band) following \cite{2010ApJ...716..348B}. We applied a $K$-correction to the 
sources following the formula

\begin{equation}
  L_{X}=4\pi d_L^2 \frac{F_{X}}{(1+z)^{2-\Gamma_{X}}},
\end{equation}

where $d_{L}$ is the luminosity distance, $F_X$ is the X-ray hard band flux, and $\Gamma_{X}$=1.7 is the photon spectral index. 
Typical uncertainties on the single luminosity do not exceed 3\% over the whole redshift range.

Figure 3 presents the hardness ratio ($HR$) distribution of the sources over the full redshift range (0.2$<z<$1.0), with the subset of radio sources 
highlighted in green.
The $HR$ was calculated as $HR=(H-S)/(H+S)$, where $S$ and $H$ denote the count rate (cts/s) 
in the soft  and hard band, respectively. As in \cite{2013A&A...557A..81M}, we refer to
AGN$_{hard}$ (obscured AGN) when $HR \geq -0.2$ and AGN$_{soft}$ (unobscured AGN) when $HR < -0.2$. This value for 
separating type 1 and type 2 AGN is a compromise based on previous studies. In particular, Garcet et al. (2007) compared optical and X-ray spectra of AGN in the 
XMM-LSS\footnote{XMM-LSS is included in the XXL-North  (XXL-N) field with
central coordinates RA=02$^h$20, DEC=-5$^d$00.} field. The authors demonstrated
that 69\% of the sources with $HR>$-0.2 are  
optical type 2 AGN and 81\% are X-ray type 2 AGN. 
Similarly, \cite{2010ApJ...716..348B} showed that 80\% of the sources with $HR \geq -0.2$ are X-ray/optically obscured AGN. 
Based on the analysis of X-ray spectra of the XXL-1000-AGN sample, 
\cite{2016A&A...592A...5F}, it was shown that higher values of $HR$ correspond to higher absorption systems (Figure 6a). 
Therefore, we first removed GAL (see above) from the AGN sample and then we divided the sample into AGN$_{soft}$ and AGN$_{hard}$ using the $HR$ criterion. 

{\bf RAD and nRAD}. We correlated the positions of the XXL-S 2.1 GHz detected sources from the ATCA catalogue of \cite{XXL Paper XVIII} with 
the positions of the optical counterparts of the X-ray point-like sources located within 1.0 arcsec. According to the probability function shown 
in Butler et al. (XXL Survey XXXI),
at 1$"$, there are {$\sim$}85\% genuine matches and {$\sim$}15\% spurious matches.
We found 270 common sources, i.e. 5.5\%  of the full sample. 
We separated the sample into two subsamples: radio (RAD) and non-radio (nRAD) sources, i.e. with and without radio counterparts, respectively.

The mean $HR$ values in the interquartile range for the samples of RAD and nRAD are
-0.63$^{+0.27}_{-0.37}$ and -0.49$^{+0.33}_{-0.51}$, respectively.

{\bf Faint and Luminous AGN}. We only considered sources detected in the hard band and we 
re-arranged $L_{X,hard}$ values in ascending order. We defined the lowest 1/4 as the `Faint AGN' sample and the highest 1/4 as the `Luminous AGN' sample. 
In the 0.2$<z<$0.6 and 0.6$<z<$1.0 redshift ranges, the log$L_{X,hard}$ (erg/s) values of the first (third) quartiles are 42.86 (43.42) and 43.57 (44.04).
Taking into account typical errors for the luminosity (see Eq. 1), the samples of faint and luminous AGN do not overlap.

\begin{figure}
\includegraphics[width=8cm,trim=0 0 5 5,clip]{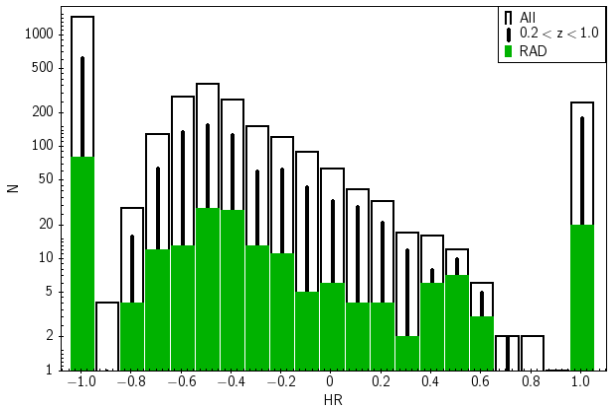}
\caption{Hardness ratio ($HR$) distribution of the XXL-S X-ray point-like sources with spectroscopic (N=3280) redshifts and those in the 0.2$<z<$1.0 redshift range.
The green histogram represents the $HR$ distribution of the radio-selected X-ray point-like sources (RAD; N=270).}
\label{3}
\end{figure}

We note that the selection function of the X-ray sources has not been considered in this paper.  We have to presume that
a number of obscured (hard) sources is not detected within the current flux limit of the survey, even though of the same intrinsic
luminosities with detected unobscured (soft) sources. Consequently, at a fixed flux limit the obscured sources are
intrinsically more luminous than the unobscured ones. This may introduce a bias to the results, which is partially taken into account by using the random sampling 
method described in detail in the next section.

\section{Large-scale environment: the nearest neighbour distance method}

In this section we study the large-scale environment ($>$ 1 Mpc\footnote{We also have a negligible number of nearest neighbours $<$ 1 Mpc; see Figure
 4.}) 
of X-ray point-like sources using the nearest neighbour method, 
assuming that in general they closely trace the distribution of optical galaxies.

It was mentioned in the previous section that we chose to analyse the distribution of X-ray point-like  sources in two redshift ranges: $0.2<z<0.6$ and $0.6<z<1.0$. 
We excluded from this analysis 274 objects with $z<$0.2 since this sample mainly comprises nearby faint galaxies (79\%). 
In order to use consistent samples for the large- and the small-scale analysis, we discarded all sources with $z>$1 
as well. For all these high-redshift objects, we cannot study the fainter environment because of the completeness limit of 
the optical catalogue (see next section for details).

We located the nearest neighbour of each X-ray source from the XXL-S catalogue and calculated the
corresponding value of the nearest neighbour distance ($d_{NNb}$, Mpc) using comoving coordinates and %In  \S3.1 
compared the mean values of $d_{NNb}$ for each pair of sources, 
i.e. GAL versus AGN, AGN$_{hard}$ versus AGN$_{soft}$, RAD versus nRAD, and faint versus luminous AGN.

\begin{figure}
\epsfig{file=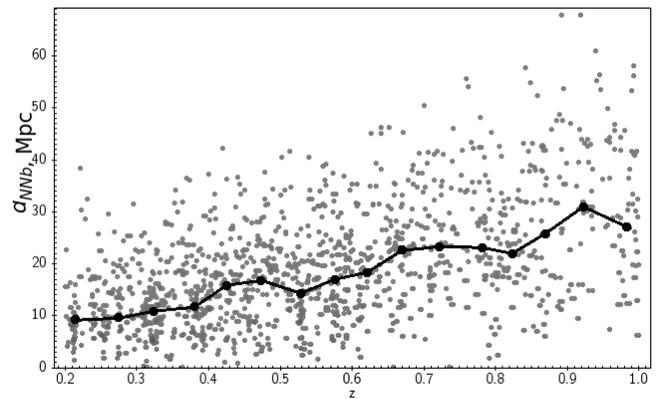,width=8.5 cm} 
\caption{Nearest neighbour distance vs. redshift for the considered sample. The broken line shows the median values of $d_{NNb}$ in each redshift bin.}
\label{4}
\end{figure}

Figure 4 illustrates the dependence of $d_{NNb}$ versus the redshift. 
Due to the lack of fainter sources at high redshifts (i.e. because of the selection function), 
the value of $d_{NNb}$ increases with redshift. 
To perform a more consistent analysis, we studied the properties of the selected sources within the two above-mentioned redshift ranges.

\begin{figure*}
\begin{tabular}{cc}
 \epsfig{file=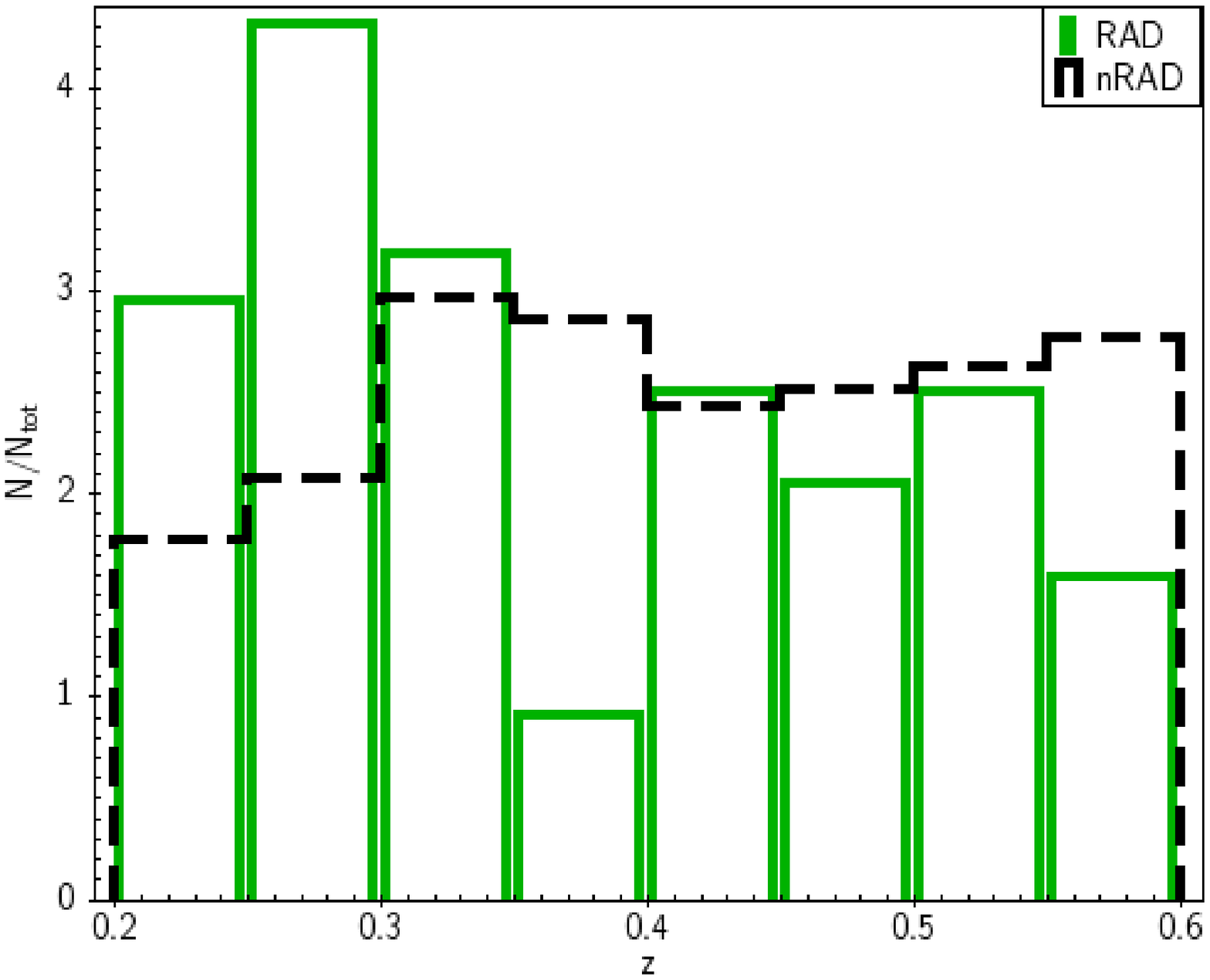,trim=0 0 10 5,clip,width=8.5cm} &
 \epsfig{file=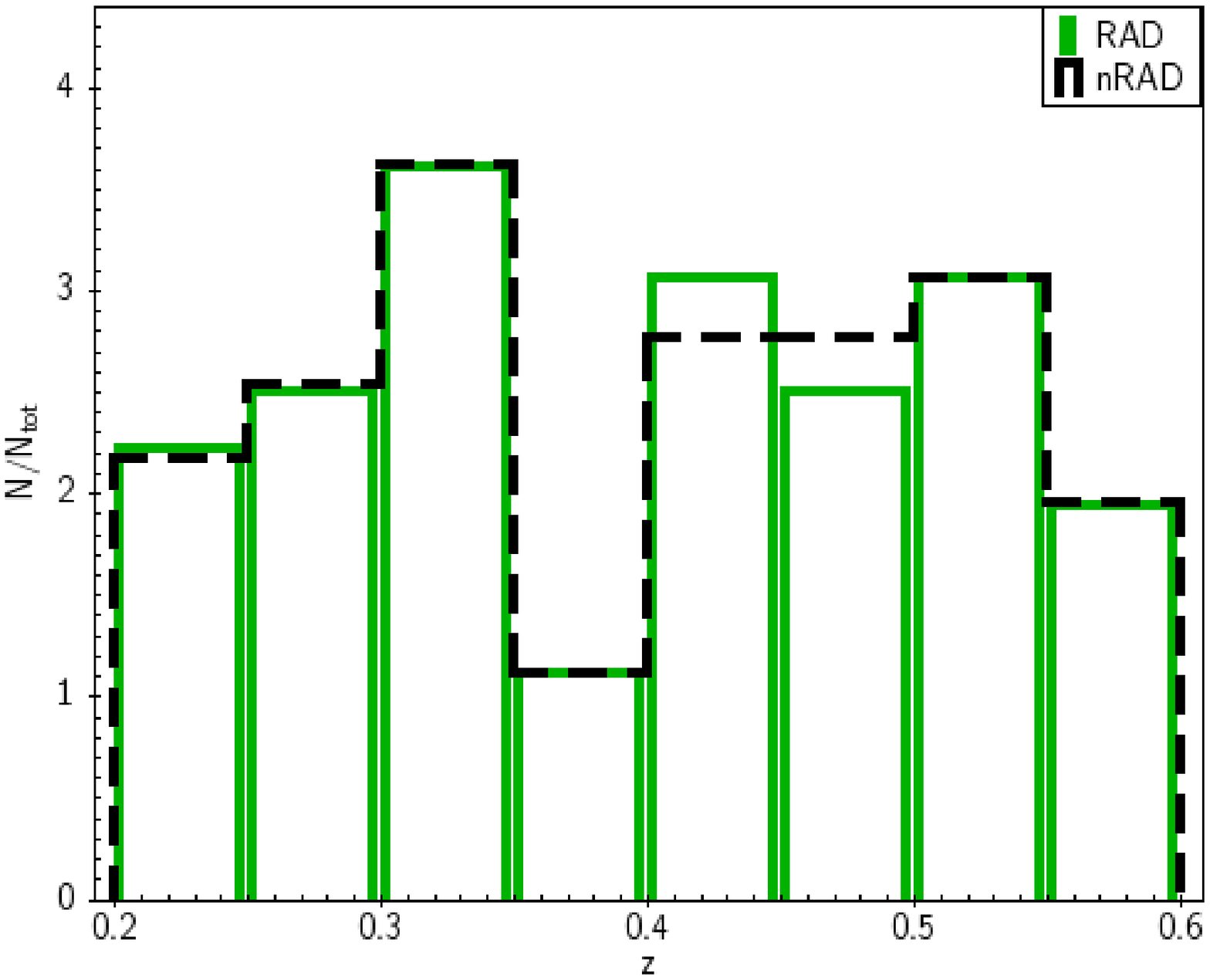,trim=0 0 10 5,clip,width=8.5 cm} \\

 \end{tabular}
\caption{Left panel: Redshift distribution of the full radio (RAD) and non-radio (nRAD) samples. Right panel: Respective randomized samples (single realization).}
\label{5}
\end{figure*}

Additionally, to eliminate any possible biases caused by the different 
redshift distributions of the compared pair samples, we applied the stratified random sampling method.
In more detail, we divided the number of sources in bins with step $\delta z$=0.05. If the number of sources in one sample was larger than in the other for
a certain bin, 
we randomly discarded N number of sources from the larger sample so as to match the distributions. Applying this procedure to all bins, 
we obtained the new normalized redshift distributions, which are not significantly different according to the Kolmogorov--Smirnov test 
(the null hypothesis that the two samples are drawn from the same parent population cannot be rejected at any significant statistical level,
see Table 2). Figure 5 illustrates this approach for the RAD/nRAD samples comparing the real (initial) distribution (left panel) 
and new randomized distribution (right panel). 

\begin{table}
\caption{Comparison of the median $z$, $HR$, and $d_{NNb}$ values for the paired subsamples. (1) subsample name; 
(2) number of members; (3) median values of the mean $z$, (5) $HR$, and (7) $d_{NNb}$, Mpc of ten randomized samples
and the respective Kolmogorov--Smirnov two-sample test probabilities of the distributions (4, 6, 8).}
\tabcolsep 3.5 pt

\begin{tabular}{|l|l|l|c|l|c|l|c|} \hline

Samples & N & $z$& $p_{KS,z_{1,2}}$ & $HR$& $p_{KS,HR_{1,2}}$ & $d_{NNb}$& $p_{KS,d_{NNb1,2}}$ \\ 
(1a)   & (2a)  & (3a)  & (4) & (5a)  & (6) & (7a)  & (8) \\ 
(1b)    & (2b)   & (3b)   &     & (5b)   &     & (7b) & \\ \hline
 \multicolumn{8}{c}{$0.2<z<0.6$}  \\    \hline        
AGN$_{hard}$  & 237     &       0.41    &       0.99    &       0.56    &       -       & 14.28   &       0.63    \\
AGN$_{soft}$   &  537 &  0.42 & & -0.74 & & 14.22 & \\ \hline
GAL  &  42      &       0.31    &       0.96    & -0.94 &       $10^{-15}$      &       11.15   &       0.80    \\
AGN$^{*}$ & 349 &  0.31 &  & -0.25 & &  11.97 & \\ \hline
RAD     & 72 &  0.40    &       0.98    &       -0.19   &       0.01    &       13.84   &       0.46    \\
nRAD &  757 & 0.40 &  &  -0.40 & &  13.61 & \\ \hline
Faint &  43 & 0.37  & 0.68  &  0.23     & $10^{-5}$ &  12.83 &  0.86    \\ 
Lum.AGN & 41 & 0.38 &  &  -0.31 & &  13.52 & \\ \hline
 \multicolumn{8}{c}{$0.6<z<1.0$}  \\    \hline    
AGN$_{hard}$    &       83      &       0.77    &       0.99 & 0.36     &       - &       24.68   &       0.66    \\
AGN$_{soft}$ &  430 & 0.77 & & -0.72 & & 24.18 & \\ \hline
RAD     &       20      &       0.78    &       0.99    &       -0.64   &       0.24    &       24.38   &       0.76    \\
nRAD & 341 & 0.78 & & -0.72 & & 23.95 & \\ \hline
Faint & 33 & 0.75  & 0.65  & -0.11      & 0.03  & 22.18 &       0.29    \\ 
Lum.AGN & 33  & 0.76 & & -0.40 & & 25.50 & \\ \hline
\end{tabular}

$^{*}$Comparison of GAL and AGN was made only in the 0.2$<z<$0.45 redshift range (limit of the GAL sample).

\end{table}

Using a single random sampling might cause an overestimation or underestimation of the real values due to a possible contamination of outliers, especially
if the number of counts is small. Therefore, 
we performed the stratified random sampling technique ten times (cross validation technique), for each redshift range. In Table 2 we report the median values (of the ten
realizations) for the redshift $z$, the hardness ratio $HR$, and the
nearest neighbour distance $d_{NNb}$.
Also, we list the two-sample Kolmogorov--Smirnov results $p_{KS}$\footnote{We also computed the t-test probabilities comparing the mean values of all corresponding pairs of parameters. 
Since we reach the same conclusions, we decided not to present them in the tables.} for the $z$, $HR$, and $d_{NNb}$ distributions of the pairs. 
Comparing the corresponding values of $d_{NNb}$, we conclude that there is no significant difference between 
the paired distributions. Therefore, the large-scale environment does not  noticeably influence the hardness or the radio activity of the sources, the presence or not of an AGN, and 
the luminosity of an AGN. However, the objects in the  faint sample are far more obscured than their luminous peers. 

A significant difference in the $HR$ distributions is also evident between the GAL/AGN, and RAD/nRAD subsamples within the low-redshift range. 
The reason for the difference between galaxies and AGN is obvious; only the AGN torus can provide this amount of obscuration.
The difference between the radio and non-radio sources cannot be readily explained. It implies, however, that the radio sources are more obscured 
than non-radio X-ray selected sources (although not at the level of the faint subsample).

Therefore, we did not find any significant differences between the populations. 
We conclude that the large-scale environment does not  significantly affect the considered X-ray and radio properties of the sources (see further discussion in Section 6.1).

\section{Small-scale environment}

The reason to study the small-scale environment of the X-ray point-like sources is that we can possibly detect variations (which are not traceable in the large-scale analysis) that may play a role in the type of activity (see e.g. Koulouridis et al. 2006).
To this end, we calculated the local (0-0.4 Mpc) optical galaxy overdensities around the sources of our subsamples. 
Due to the lack of extensive spectroscopic coverage of the normal galaxy population (non X-ray sources) we performed a projected (2D) overdensity analysis, 
following \cite{2013A&A...557A..81M}, where we used a similar approach for the XMM-LSS field. We considered the same subsamples and redshift ranges as in the previous sections.

\subsection{Methodology}
Taking into account the redshift and the angular
distance $D_{A}$, we estimated the angular sizes of the linear radius 0.4 Mpc at the
source rest-frame.  
Then we counted the number, $N_{0}$, of the BCS optical galaxies 
within a circle of radius 0.4 Mpc and within a range of magnitudes from $m^{*}-\Delta m$ to
$m^{*}$ (bright environment) and from $m^{*}$ to $m^{*}+\Delta m$ (faint environment), 
where $\Delta m = 1$ and $m^{*}$ is the apparent magnitude corresponding to the knee of the $i'$-band luminosity function [$\Phi(L)$], given by
\begin{equation}
  m^{*}=5\log_{10} d_{L} + 25 +M_{i'}^{*}+Q_{0.1}(z)+K_{0.1}(z) \;,
\end{equation}
where $M_{i'}^{*} (=-20.82+5\log_{10}h)$ is the absolute magnitude at the knee of
the $i$-band $\Phi(L)$ taken from \cite{2003ApJ...592..819B};
$Q(z)$ and $K(z)$ are the evolution and $K$-corrections, respectively, taken
from \cite{1997A&AS..122..399P} and shifted to match their rest-frame
shape at $z=0.1$; and $d_{L}$ is the luminosity distance. Here, we used the magnitudes as a proxy for the stellar mass. 
The $\Delta m$ limits are similar to what is used to evaluate the cluster members (e.g. Martini et al. 2013, Koulouridis et al. 2014, and  Bufanda et al. 2017).

Next we calculated the galaxy overdensities, $\delta$,
\begin{equation}
\delta=\frac{N_{0}A_{b}}{N_{b}A_{0}}-1 \;,
\end{equation}
where $N_{0}$ is the total number of objects in considered circle with surface area $A_{0}$, and 
$N_{b}$ is the local background counts estimated in the annulus between 3.1 and 5 Mpc with surface area
$A_{b}$.  

To estimate the $\sigma$ uncertainties we used the Jackknife resampling technique (Efron et al. 1983). To check the significance of the results we
compared the overdensity of galaxies around each sources with that expected from the
mock X-ray source distributions. The mock catalogues have random coordinates, but
the same fiducial magnitude ($m^{*}$) distribution. For the mock randomly distributed sources, we
used the same BCS optical catalogue to search for neighbours as we did for the real ones including the Jackknife technique.

  For each catalogue we calculated the cumulative overdensity distribution 
  
  \begin{equation}
  F(>\delta)=N(>\delta)/N_{tot},
  \end{equation}
  
which is defined as the percentage of all sources $N_{tot}$ having an overdensity above a given $\delta$ value.
  Finally,  we compared the real distributions with the random ones.

  \subsection{Results}
    
  First, we calculated the overdensity distributions for the different randomized subsamples in two redshift ranges for the bright 
  ($m^{*}-1 < m< m^{*}$) and faint ($m^{*}< m< m^{*}+1$) environments and compared them with the mock distributions. 
  We did not study the faint environment of sources in the high-redshift range because of the completeness of the optical BCS catalogue at about $i\sim$ 23, 
  which corresponds to our rest frame m* at $z$=1. 
  
  We used the Kolmogorov--Smirnov (KS) two-sample test to estimate quantitatively the differences between the
  real and mock overdensity distributions. The typical KS probabilities ($p_{KS}$) 
  of the two corresponding distributions of overdensities being drawn from the same parent population are $<10^{-3}$, 
  meaning that the real and mock distributions are significantly different in the first redshift range. The exceptions are the GAL overdensities in both environments and
  the luminous and faint AGN in bright environments and in both redshift ranges.
   
  Figures 6 and 7 present the cumulative distributions of overdensities for the real and mock 
  subsamples in the low- and high-redshift range, respectively. 
   In Table 3 we list the value of $F(>\delta)$ at the point $\delta>0$. We also list the corresponding $1\sigma$ intervals for the real and mock samples. 
   The $\sigma_s$ indicates the difference (in $\sigma$ units) that the fraction of real sources at a 
   given overdensity level is higher than the corresponding fraction of the mock sources. For example, in the $0.2< z <0.6$ redshift range, the fraction of AGN$_{hard}$ 
   in bright overdense environments with $\delta>0$ is 66\%. This is higher
  than the random expectations at the 5.6$\sigma$ level. In the faint overdense environment, 
  the corresponding fraction is 65\%, which is higher than the random expectations at the 4.0$\sigma$ level. 
  In the high-redshift range, we find $F(\delta>0)= 55$\%, which is not significantly different from the random expectations at the 0.3$\sigma$ level.

 Our results show that, in general, X-ray point-like sources inhabit both dense and
 underdense environments. However, there are significantly more X-ray sources than in their corresponding mock catalogues 
 which inhabit overdense regions in both redshift ranges (in agreement with Melnyk et al. 2013), especially in the low-redshift range. 
 The difference is less than $3\sigma$ only for GAL, RAD, faint, and luminous AGN  in both environments. However, the low significance is most likely due to small sample sizes. 
 At higher redshifts we do not have clear evidence of overdensities (except for the nRAD sample) probably because of the  small  sample sizes.

  These results are not unexpected given that AGN are known to be clustered. Our main interest, though, is 
  to study the environmental trends for the subsamples defined in Table 1.
  
  In the last column of Table 3 ($\sigma_{p}$) we compare the real samples at the point of $\delta$=0 in $\sigma$ units. 
  As can be seen, there are no significant differences for any pair of samples. 
  Therefore, pairs of GAL and AGN, AGN$_{hard}$ and AGN$_{soft}$, radio and non-radio sources, and faint and luminous AGN occupy similar environments.

It is worth mentioning that the small- and large-scale environments of our sources do not correlate, 
i.e. isolated objects as defined in the large-scale analysis are found in a variety of small-scale environments, not necessarily in isolation.

\begin{figure*}
\begin{tabular}{cc}
\epsfig{file=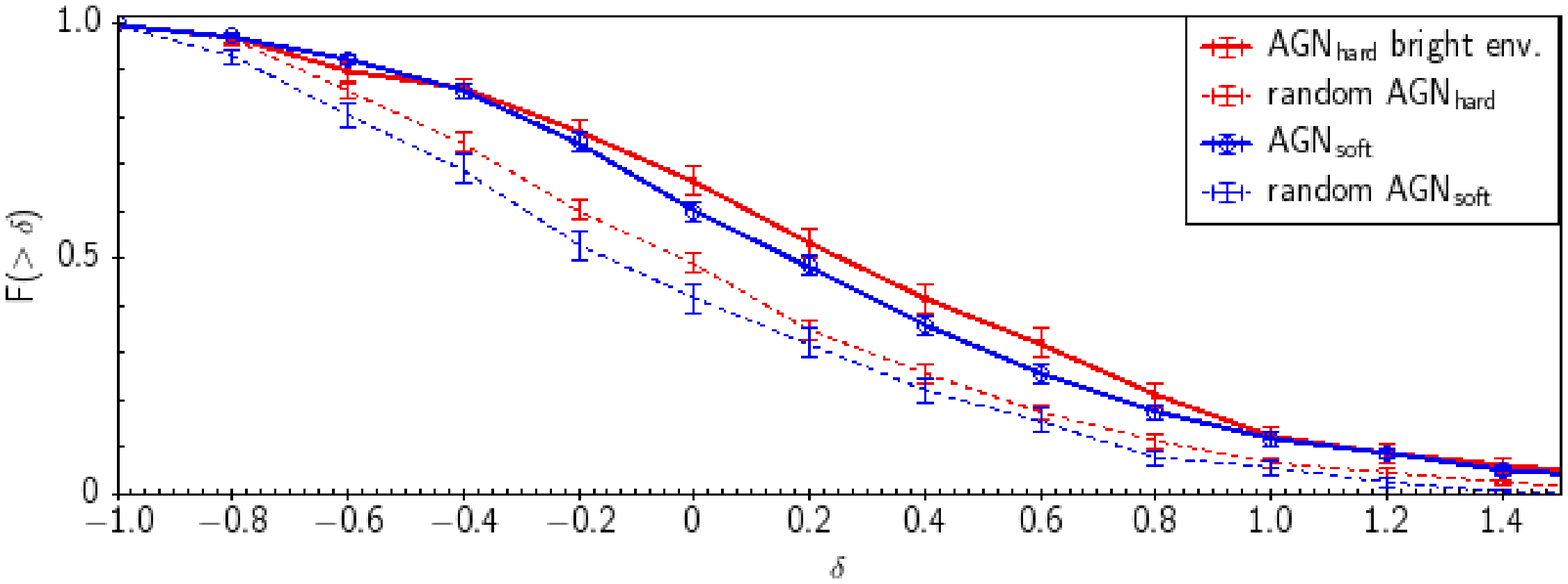,width=8.5 cm} &
\epsfig{file=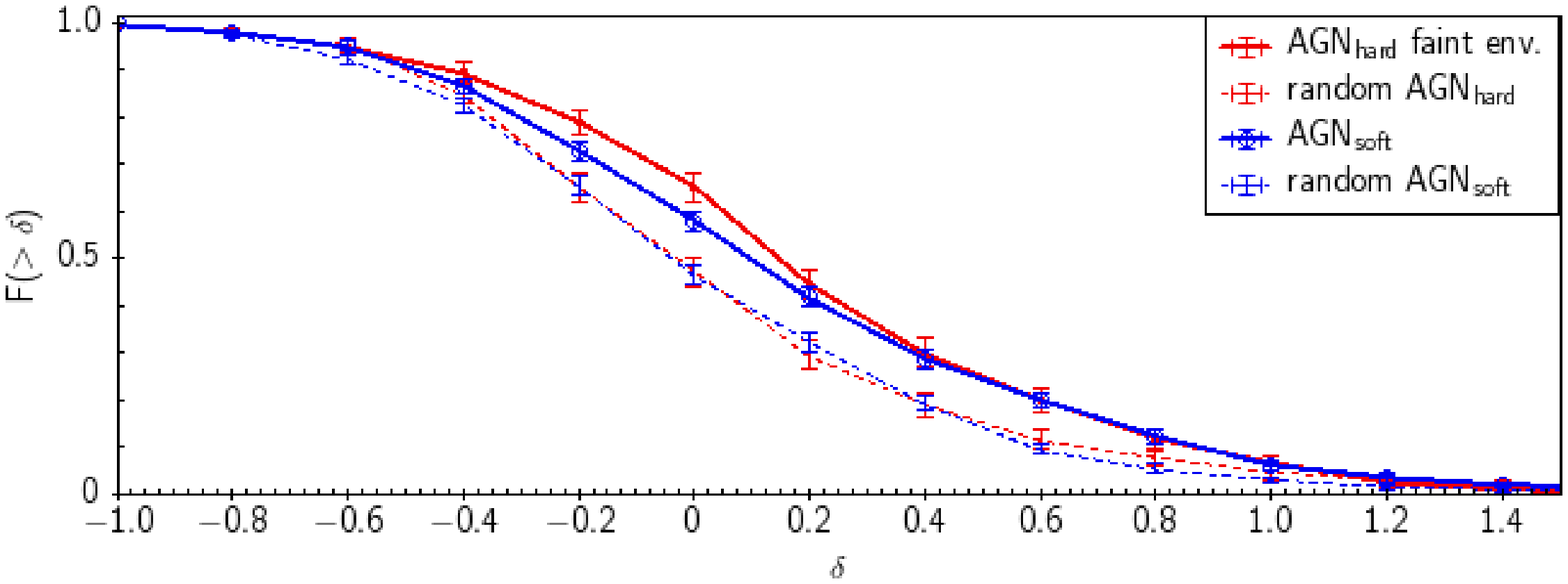,width=8.5 cm} \\
\epsfig{file=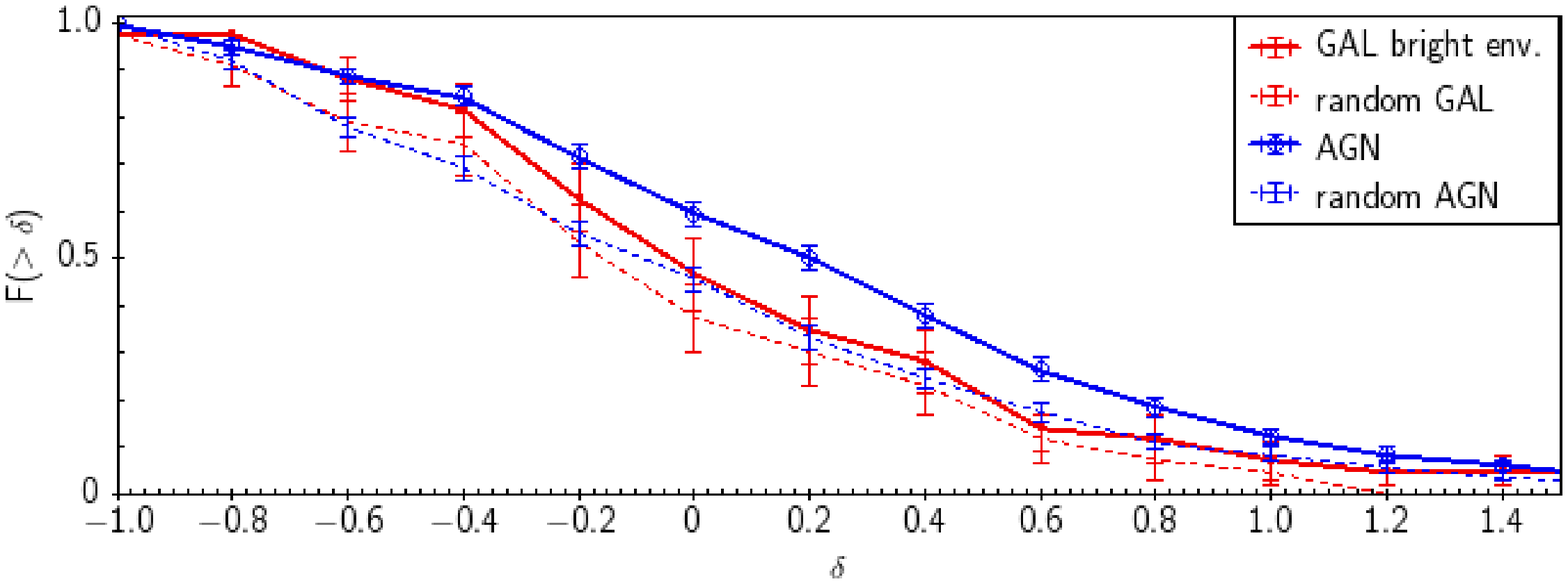,width=8.5 cm} &
\epsfig{file=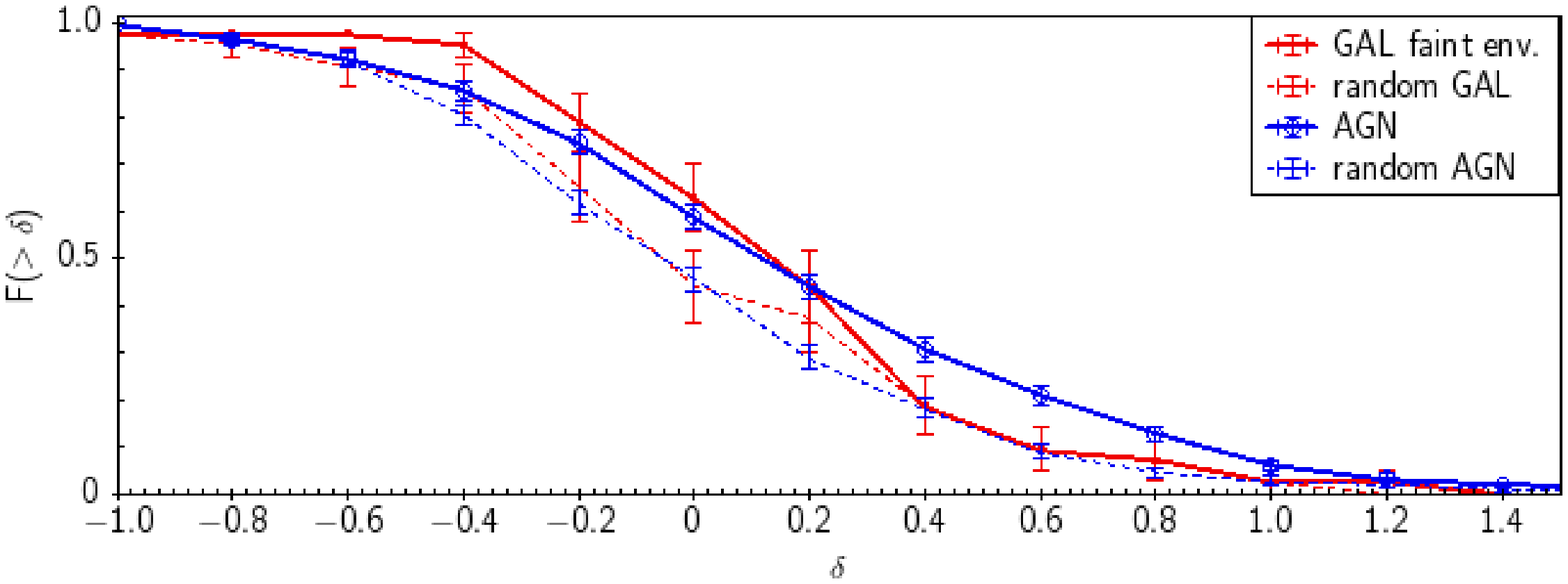,width=8.5 cm} \\
\epsfig{file=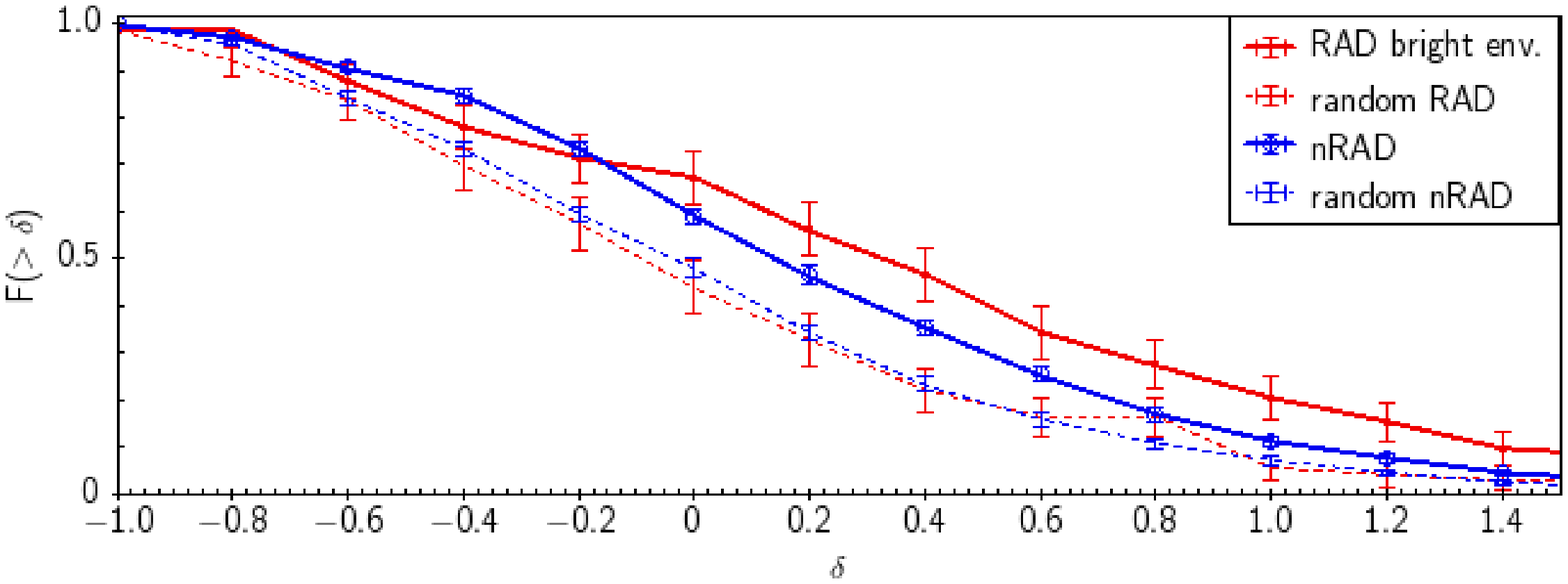,width=8.5 cm} &
\epsfig{file=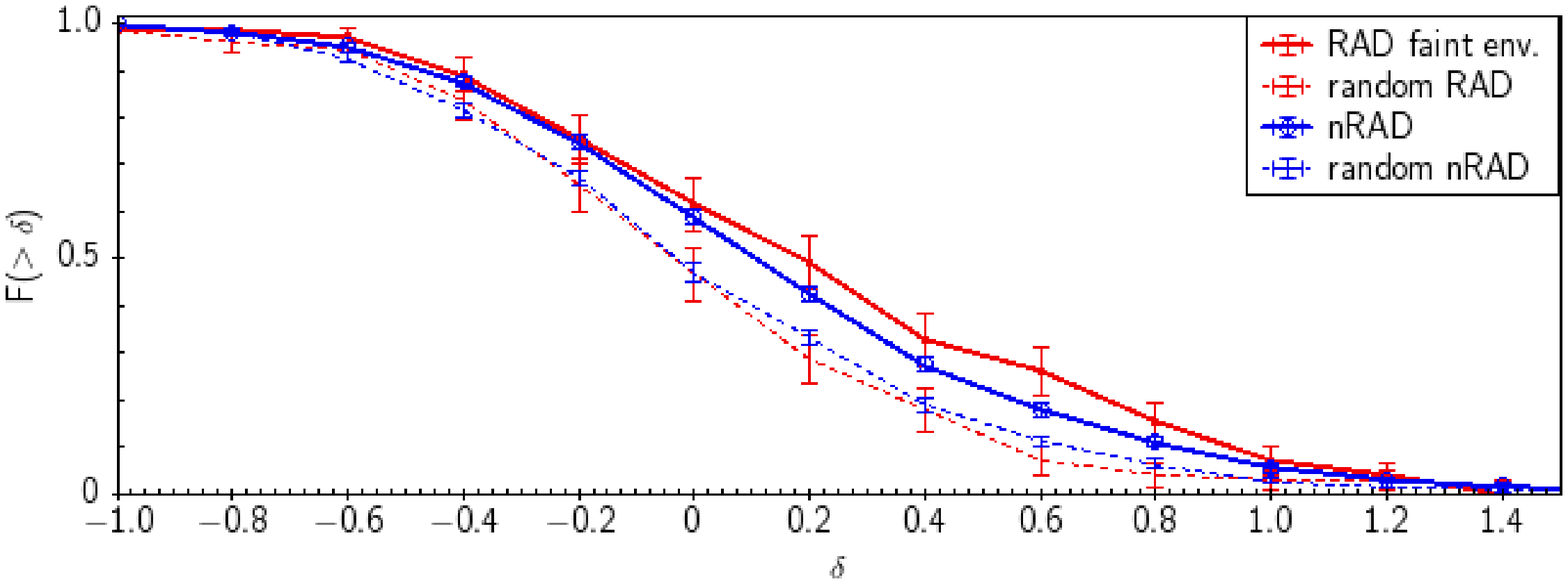,width=8.5 cm} \\
\epsfig{file=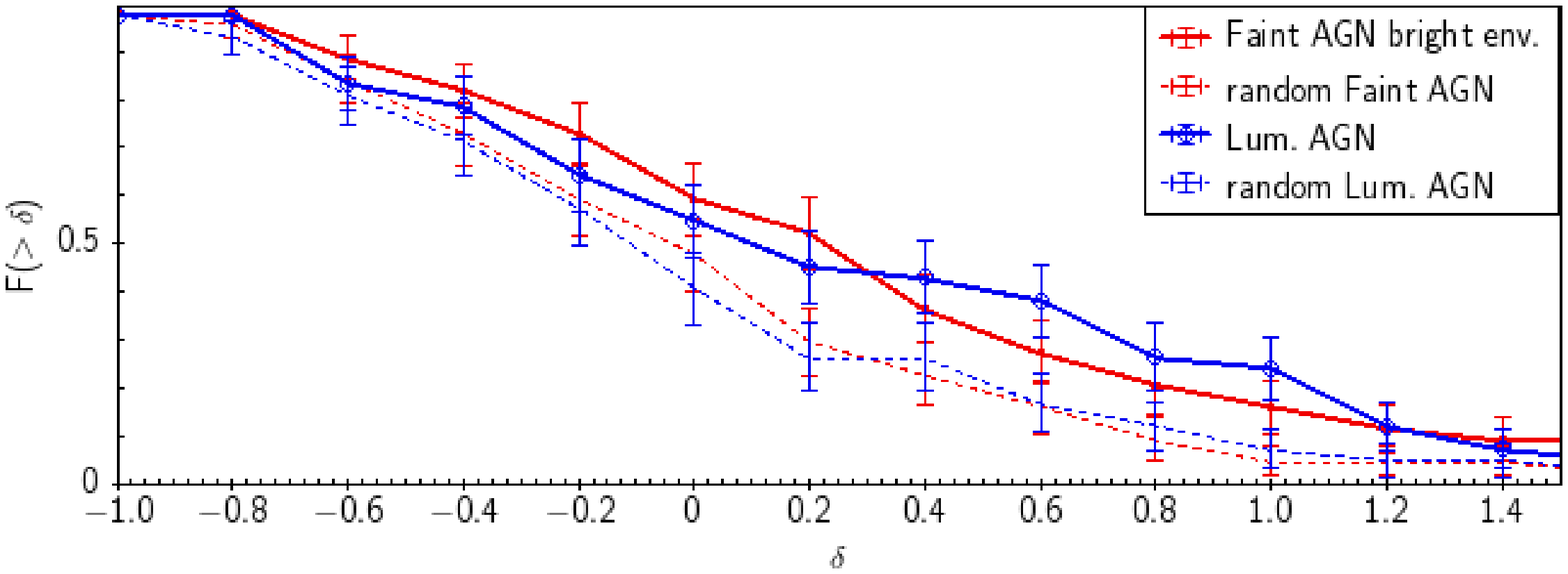,width=8.5 cm} &
\epsfig{file=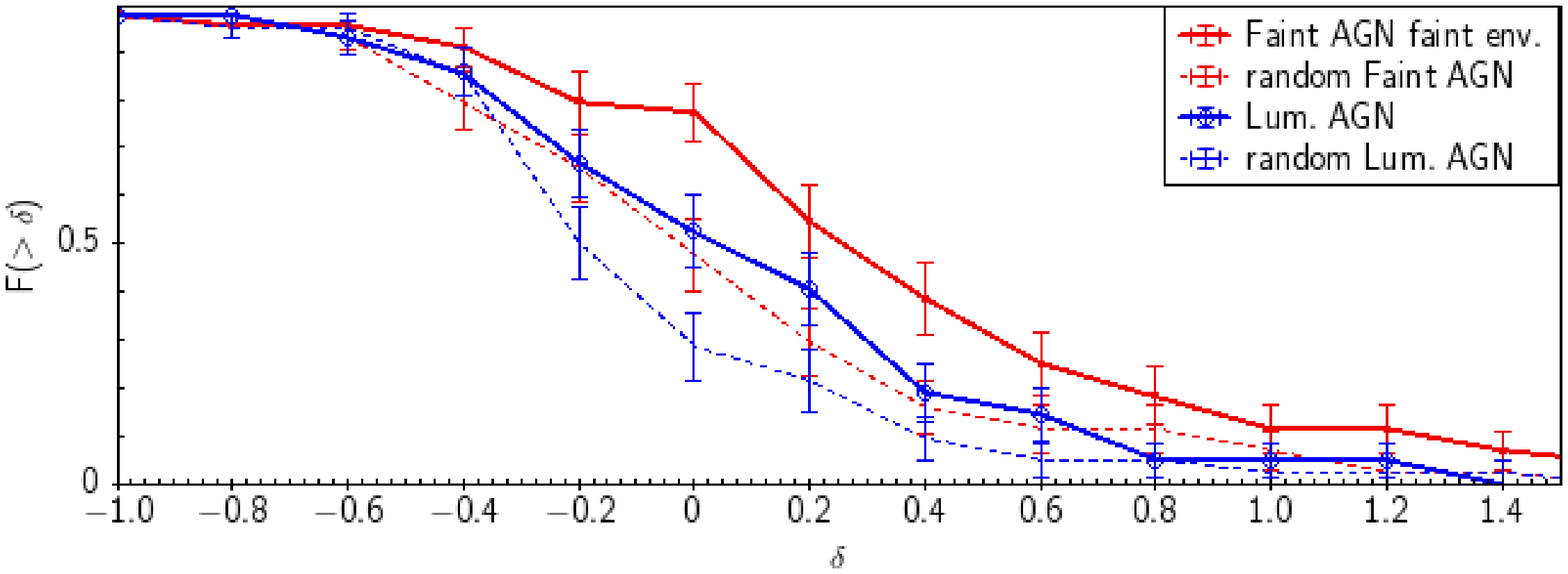,width=8.5 cm} \\
 \end{tabular}
\caption{Cumulative distributions of the overdensities for the
    randomized and mock samples in the  $0.2 < z < 0.6$ redshift bin. The different samples are coded as indicated in the label of each panel. 
    The error bars correspond to 1 $\sigma$ deviations.}
\label{6}
\end{figure*}

\begin{figure}
\begin{tabular}{c}
\epsfig{file=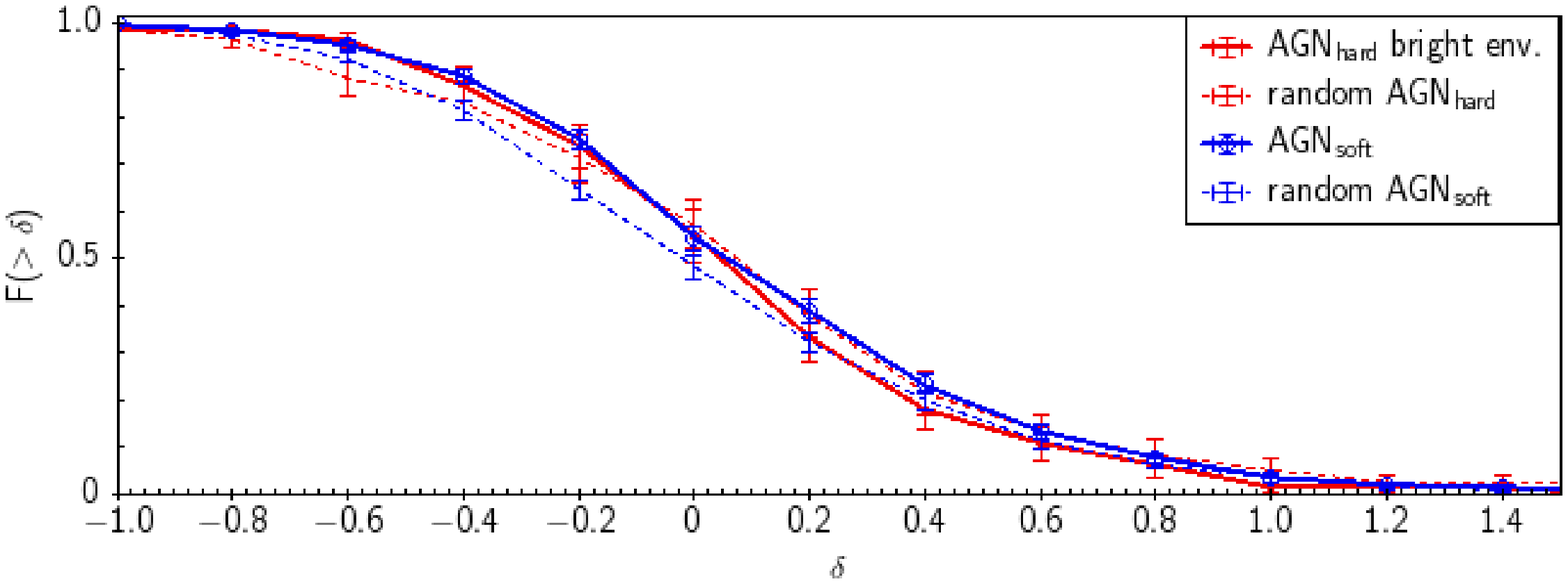,width=8.5 cm} \\
\epsfig{file=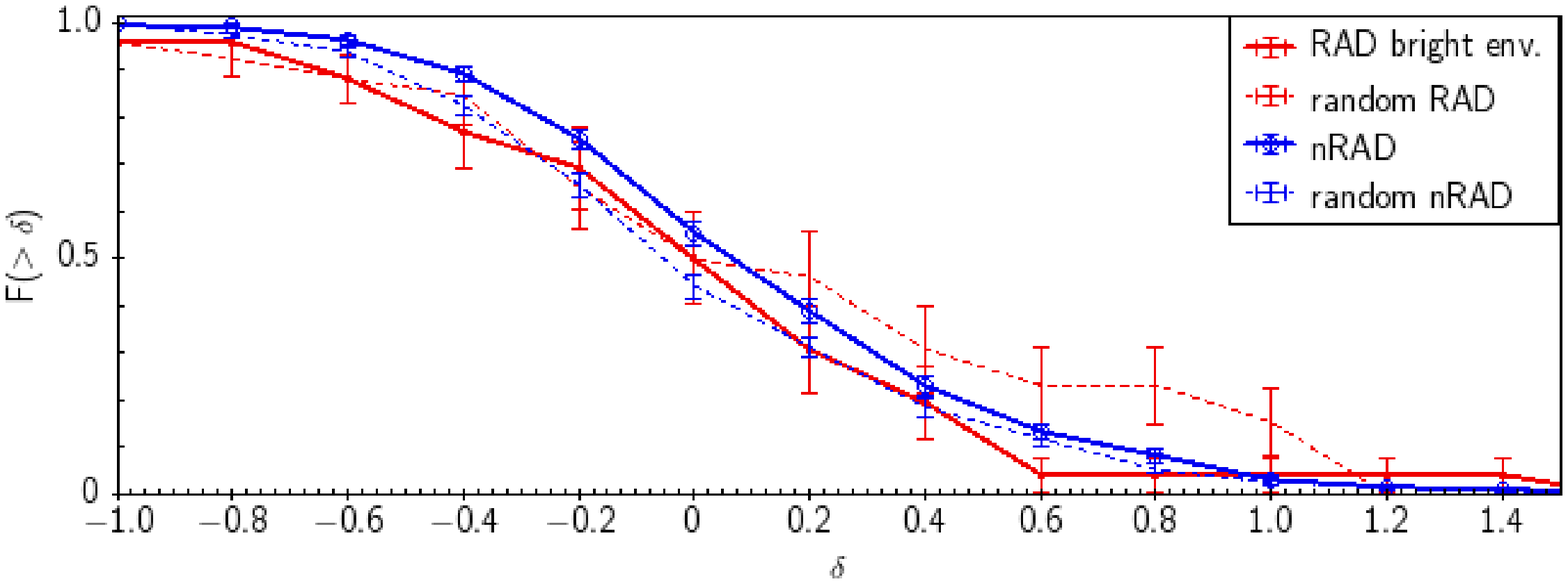,width=8.5 cm} \\
\epsfig{file=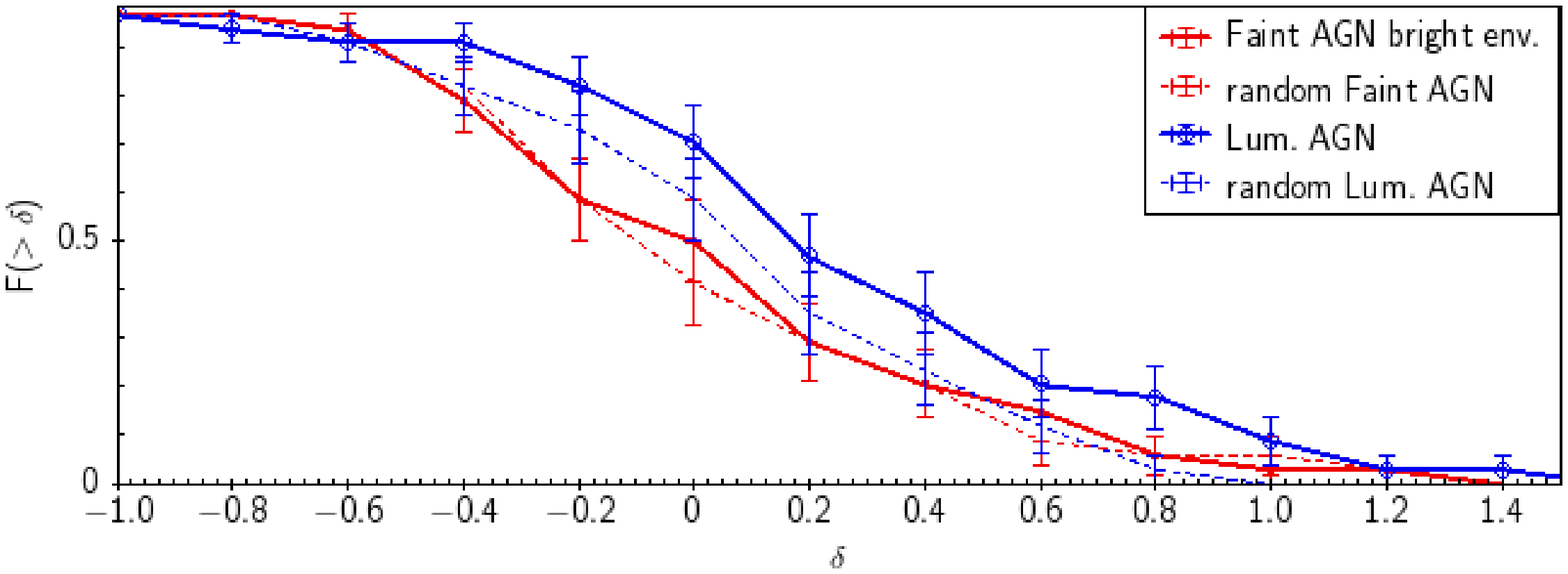,width=8.5 cm} \\

 \end{tabular}
\caption{Cumulative distributions of the overdensities for the
    randomized and mock samples in the  $0.6 < z < 1.0$ redshift bin. The different samples are coded as indicated in the label of each panel. 
    The error bars correspond to 1$\sigma$ deviations.}
\label{7}
\end{figure}

\begin{table}
\caption{Comparison of the fraction $F$ of sources with overdensity values  $\delta>0$  in the randomized samples and their respective mock catalogues.  
$\sigma_s$ is the difference between $F_{real}$ and $F_{mock}$ in units of its typical standard deviation, calculated as
$|F_{real}-F_{mock}| /(\sigma_{real}^2+\sigma_{rand}^2)^{1/2}$;  $\sigma_p$ is the corresponding difference between pairs of the samples AGN$_{hard}$/AGN$_{soft}$, etc.}
\tabcolsep 4 pt

\begin{tabular}{|l|l|c|c|c|c|} \hline

Sample  & $N$   &       $F_{real(>\delta)}\pm \sigma_{real}$,\% & $F_{mock(>\delta)}\pm \sigma_{mock}$,\%       & $\sigma_s$ & $\sigma_p$ \\ \hline
 \multicolumn{5}{c}{$0.2<z<0.6$, bright environment}  \\
\hline
 AGN$_{hard}$   & 237   & 66$\pm$3.0 &  42$\pm$3.2 & 5.6 & 1.7 \\                                               
AGN$_{soft}$    & 537   & 60$\pm$2.1 & 49$\pm$2.2 & 3.7 & \\ \hline                                                                                                     
GAL     & 42    & 46$\pm$7.6    & 37$\pm$7.4 & 0.9 & 1.6 \\
AGN     & 349   & 59$\pm$2.6 & 45$\pm$3.7 & 3.8 & \\ \hline                                                             
RAD     & 72 &  66$\pm$5.5 & 44$\pm$5.8 & 2.9 & 1.4\\                                                           
nRAD    & 757   & 59$\pm$1.8 & 48$\pm$1.8 &     4.3 & \\ \hline                                                         
Faint AGN &     43 &    59$\pm$7.4 & 48$\pm$7.5 & 1.3 & 0.4 \\
Lum. AGN & 41 & 55$\pm$7.7 & 40$\pm$7.6 & 1.3 & \\  \hline
 \multicolumn{5}{c}{$0.2<z<0.6$, faint environment}  \\  \hline
AGN$_{hard}$    & 237      &  65$\pm$3.1  &  47$\pm$3.2 & 4.0 & 2.0 \\  
AGN$_{soft}$    & 537 & 58$\pm$2.1 &    46$\pm$2.2 &  3.8 & \\ \hline
GAL     & 42    & 63$\pm$7.3 &  44$\pm$7.6 & 1.8 & 0.5 \\
AGN     & 349   & 59$\pm$2.6 & 46$\pm$2.7 & 3.4 & \\ \hline
RAD     & 72 & 62$\pm$5.7 &  47$\pm$5.8 & 1.9 & 0.5 \\
nRAD    & 757   & 59$\pm$1.8   &  47$\pm$1.8 & 4.7 & \\ \hline
Faint AGN & 43 & 77$\pm$6.1 & 48$\pm$7.5 & 3.0  & 2.5 \\
Lum. AGN & 41 & 52$\pm$7.7 & 37$\pm$7.4 & 1.6 & \\ \hline
\multicolumn{5}{c}{$0.6<z<1.0$, bright environment}   \\                                 \hline                                                                                          
AGN$_{hard}$ & 83 &     55$\pm$5.4 &    57$\pm$5.4 & 0.3 &   0.1\\
AGN$_{soft}$  & 430     & 54$\pm$2.4 &  48$\pm$2.4 & 1.8 & \\ \hline
RAD & 20 & 50$\pm$9.8 & 50$\pm$9.8      & 0.0 &  0.5 \\
nRAD & 341 & 55$\pm$2.5 & 44$\pm$2.5    & 3.2 & \\ \hline
Faint AGN & 33 & 50$\pm$8.6  & 41$\pm$8.5 & 0.7 & 1.8 \\
Lum. AGN & 33  & 71$\pm$7.6  & 59$\pm$8.4 & 1.0 & \\ \hline
\end{tabular}

\end{table}

\section{X-ray clusters versus AGN agglomerates}

In this section we study the XXL-S AGN-supercluster candidates defined as the super structures comprising X-ray point-like sources. These structures were obtained by applying the classical bottom-up hierarchical clustering 
method\footnote{Linkage of clusters starts by considering each object as a cluster. Then, one by one, it merges all elements into a single cluster. 
Depending on the input free parameter of clustering radii, the hierarchical tree is cut dividing the sample on clustered and unclustered objects.
We used the minimum distance between clusters for linkage.} 
to the sample of X-ray point-like sources using comoving coordinates. 
To distinguish between X-ray galaxy clusters (i.e. virialized structures that are detected as X-ray extended sources)
and the structures which were obtained with the hierarchical
method, we use the term `agglomerate'. This is in analogy to the study by \cite{2012AstBu..67..353K}. 
Then, we checked if the agglomerates trace the distribution of X-ray galaxy clusters (X-ray extended sources). 

In this analysis we defined AGN agglomerates
at clustering radii of 16.76 Mpc and 28.32 Mpc in the 0.2$<z<$0.6 and 0.6$<z<$1.0 redshift ranges, respectively. 
Adopting these values, 2/3 of the sources are in agglomerates and 1/3 are isolated in our sample.  Here we followed the proportion $\sim$ 70\% of clustered and 
$\sim$ 30\% unclustered galaxies in the Local Universe (see \cite{2012AstBu..67..353K} and references therein). Figure 8 shows the dependence of the 
typical distances between the sources in the agglomerates $<D>$ versus the number of their X-ray point-like members.

The XXL X-ray cluster sample (C1 and C2 types) is described in \cite{2016A&A...592A...1P}, Pacaud et al. (2016; hereafter \cite{2016A&A...592A...2P})
and Adami et al. (XXL Paper XX).
 The considered samples of spectroscopically confirmed X-ray galaxy clusters from XXL Paper XX contain $N_{cl}$=81 and 22 clusters, 
 respectively, in the 0.2$<z<$0.6 and 0.6$<z<$1.0 redshift ranges.

To avoid biases related to inhomogeneity of the X-ray point-like sources in the field with respect to cluster positions
we have performed the following test. 
Separately for the two considered redshift ranges around each X-ray cluster within with an angular radius $r$, we counted the number of 
all point-like X-ray sources such that $m^{*}-2<m<m^{*}+1$, where $m$ is the rest-frame $i$-band magnitude of a cluster and $m^{*}$ is the apparent magnitude of an X-ray AGN/galaxy.
According to this method, we calculated  the number of point-like sources with available spectroscopic redshifts and without it. 
In Figure 9 we plotted the ratio between the numbers of X-ray point-like sources with spectro-z and the whole sample for different angular radii around the X-ray clusters. 
The error bars show the boundaries of the confidence interval at the 95\% level. As can be seen, the completeness of the spectroscopic redshifts for the 
X-ray point-like sources is 78\% and 30\%  
of the total population of sources in the 0.2$<z<$0.6 and 0.6$<z<$1.0 redshift ranges, respectively.
This percentage does not depend significantly on the distance from the X-ray clusters.  
We  conclude that the distribution of the X-ray point-like sources near the X-ray clusters is non-biased.

\begin{figure}
\includegraphics[width=8.5cm,trim=0 0 0 0,clip]{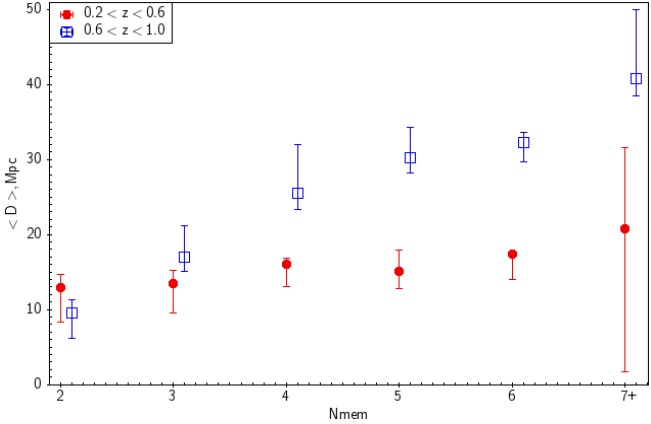}
\caption{Median values of the average distances between the members of agglomerate within their interquartile ranges vs. the number of agglomerate members.}
\label{8}
\end{figure}

\begin{figure}
\includegraphics[width=8.5cm,trim=0 0 0 0,clip]{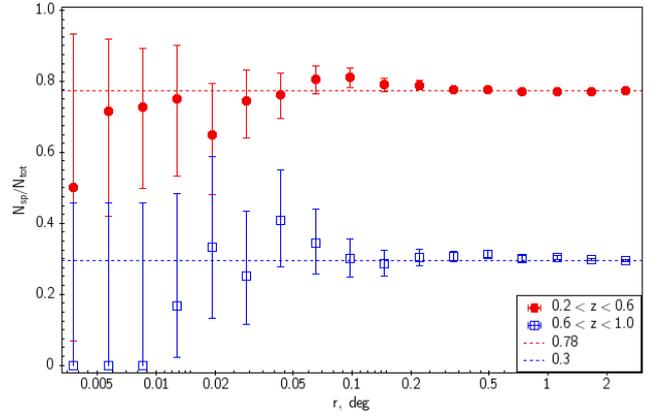}
\caption{Ratio between the numbers of X-ray point-like sources with spectroscopic redshifts (N$_{sp}$) and the whole sample (N$_{tot}$) 
for different angular radii $r$ around the X-ray clusters.}
\label{9}
\end{figure}

\begin{figure}
\includegraphics[width=8.5cm,trim=0 0 0 5,clip]{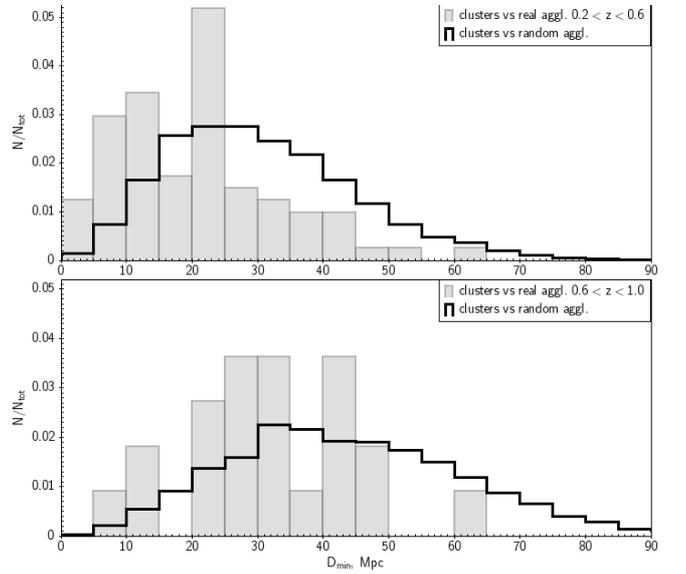}
\caption{Distributions of the minimum distances between all the X-ray clusters and all the real/mock agglomerates (2+).}
\label{10}
\end{figure}

We also computed the minimum distances between each cluster and all the agglomerates (geometrical centres of the agglomerates).
Then we built the distributions of the minimum distances and compared them with the respective distributions  
of the galaxy clusters and randomly chosen agglomerates. To build the mock catalogues, we chose random 
coordinates and redshifts from the real sample of the X-ray point-like sources in the corresponding redshift ranges. 
Therefore, our mock catalogues have the same properties as the real one. 
The number of the mock agglomerates was 100 times higher than the number of the real agglomerates.
Comparison between the minimum distance distributions
of all the X-ray clusters and the real/mock agglomerates for all the samples are shown in Figure 10. 

Table 4 presents the mean values of the minimum distances between the X-ray clusters and the real or mock AGN agglomerates with different populations: 
2+ (two or more members),  3+, 5+, and 10+. 
In both redshift ranges, the null hypothesis that the real and mock distributions are drawn from the same parent set can be rejected at a high level of significance. 
In other words, we report significantly smaller distances between the X-ray clusters and the AGN agglomerates with different a number of members than  expected from the mock 
realizations. 

Mainly, the AGN agglomerates follow the distribution of the X-ray clusters at the $<$25 Mpc scale in the 0.2$<z<$0.6 and $<$45 Mpc in the 0.6$<z<$1.0 ranges, respectively:
73$^{+7}_{-8}$\%  and 86$^{+9}_{-13}$\%. The corresponding random expectations are 39$^{+1}_{-1}$\% and 55$^{+1}_{-2}$\%, respectively. 
The above-mentioned  uncertainties correspond to the 95\% confidence level, so these results are highly significant.

In Table 5 we present the geometric centres of the most populated agglomerates ($>$10 members) and their possible associations with X-ray clusters, supercluster candidates, and 
pairs of clusters from XXL Paper XX. 
In the columns we indicate the published names of the cluster, supercluster, or pair and the 
minimum distances between the corresponding galaxy structure and the given agglomerate. We conclude that the most populated concentrations of AGN are associated with supercluster 
candidates.

\begin{table}
\caption{ Kolmogorov--Smirnov $p_{KS}$ and t-test  $p_{t}$ probabilities of the real and mock minimum distance distributions ($D_{min,real}$ and $D_{min,mock}$ in Mpc) 
and the corresponding mean values between the clusters and the agglomerates with different populations being drawn from the same parent population. 
For example, 2+ means two or more members in an agglomerate.}
\tabcolsep 2.3 pt

\begin{tabular}{|l|l|l|c|c|c|c|} \hline
Aggl. & $N_{Aggl}$      &       $N_{cl}$ & $<D_{min,real}>\pm SD$, & $<D_{min,mock}>\pm SD$ &   $p_{KS}$ & $p_t$                \\ \hline       
                        \multicolumn{7}{c}{$0.2<z<0.6$}                 \\ \hline                  
2+ & 143 &      81      & 21.0$\pm$12.3 & 30.6$\pm$14.2 &       10$^{-9}$        & 10$^{-9}$\\
3+ & 84 &       81      & 25.0$\pm$15.2 & 37.6$\pm$14.9 &       10$^{-10}$       & 10$^{-10}$\\
5+ & 39 &       81      & 32.2$\pm$20.2 & 50.1$\pm$24.9 &       10$^{-11}$       & 10$^{-11}$\\
10+ & 12 &      81      & 55.7$\pm$36.2 & 89.7$\pm$59.2 &       10$^{-10}$       & 10$^{-10}$\\ \hline
                        \multicolumn{7}{c}{$0.6<z<1.0$} \\ \hline
2+ & 104        & 22    & 32.4$\pm$13.0 & 44.3$\pm$19.2 &       10$^{-5}$ & 10$^{-5}$     \\ 
3+ & 54 & 22    & 43.2$\pm$17.8 & 57.9$\pm$26.3 &       10$^{-4}$        & 10$^{-4}$\\ 
5+ & 20 & 22 & 60.5$\pm$26.6 & 88.1$\pm$41.8 &  10$^{-6}$ & $10^{-5}$ \\ 
10+ & 6 & 22 & 109.1$\pm$75.6 & 146.7$\pm$76.0 & 10$^{-3}$ & $10^{-4}$ \\ \hline
\end{tabular}

\end{table}

\begin{table*}
\caption{XXL-S supercluster candidates,  the most populated agglomerates, and possible correlations with X-ray clusters. $N_{mem}$ represents the number of members in the 
agglomerate, RA and
DEC are the coordinates of the centre, $<z>$ is the average redshift, $D_{max}$ is the maximum distance
between members of the agglomerate, $<D>$ is the average distance between its members, and 
$D_{min,real}$ is the distance between a given cluster and its nearest agglomerate.}
\tabcolsep 8 pt
{\centering
\begin{tabular}{|l|c|c|c|c|c|c|c|c|} \hline
$N_{mem}$       &       RA      &       DEC     &       $<z>$   & $D_{max}$, Mpc &   $<D>$, Mpc &    XLSSC clusters* & $D_{min,real}$, Mpc & Supercluster**\\ \hline
46      &       352.179 &       -54.240 &       0.21    &       84.4    &       37.6    &       595, 586, 608 & 12, 13, 12 & part S05 \\     
38      &       355.082 &       -54.440 &       0.33    &       80.2    &       35.2    &       614, 548, 632, 538   & 24, 21, 21, 24  & part S02 \\
30      &       352.810 &       -53.872 &       0.26    &       86.1    &       36.1    &       ---     & ---  & --- \\
24      &       356.087 &       -53.811 &       0.62    &       135.6   &       58.2    &       509 & 40 & --- \\
21      &       352.315 &       -54.810 &       0.28    &       74.7    &       32.6    &       612, 622, 588, 519, 524 & 16, 22, 19, 22, 21 & part S03 \\
17      &       353.983 &       -55.115 &       0.38    &       72.9    &       35.5    &       573, 543     & 15, 23  & --- \\
17      &       351.601 &       -54.448 &       0.61    &       110.9   &       50.9    &       580,611 & 25, 6 & pair, id30 \\
15      &       352.953 &       -56.132 &       0.31    &       67.1    &       26.5    &       ---     & ---  & --- \\   
14      &       352.677 &       -52.489 &       0.45    &       63.5    &       30.5    &       561, 641 & 21, 24 & pair, id25       \\      
13      &       350.457 &       -53.179 &       0.71    &       90.7    &       40.7    &       ---     & ---  & --- \\           
13      &       349.393 &       -54.392 &       0.87    &       114.2   &       49.9    &       ---     & ---  & --- \\           
13      &       349.607 &       -54.196 &       0.66    &       70.65   &       37.6    &       ---     & ---  & --- \\   
11      &       349.183 &       -54.948 &       0.44    &       41.6    &       17.1    &       ---     &  --- & --- \\   
11      &       354.043 &       -54.479 &       0.53    &       46.8    &       26.1    &       ---     & ---  & --- \\                                                                                                           
11      &       350.214 &       -53.125 &       0.37    &        68.3   &       35.4    &       547         & 20  & --- \\          
10      &       349.291 &       -53.780 &       0.28    &       34.2    &       21.9    &       526, 557, 591        & 8 ,13, 17  & part S06 \\      
10      &       354.541 &       -56.039 &       0.47    &        41.7   &       20.8    &       639, 609 & 17, 8  & --- \\
10      &       349.773 &       -53.559 &       0.79    &       91.1    &       47.6    &       560     & 42  & --- \\    \hline  
\end{tabular}
}

*  The name of the X-ray cluster given in XXL Paper XX.

**  Supercluster candidates (and also cluster pair) defined in XXL Paper XX as concentrations of X-ray clusters, `part' means that  half or more of the members of 
the supercluster are associated with the corresponding agglomerate.

\end{table*}

\section{Discussion and conclusions}

We have studied in the present paper the environment and the clustering properties of the X-ray selected point-like sources with spectroscopic redshifts 
from the 25 deg$^2$ XXL-S field in  two redshift ranges: 
 1012 sources with $0.2<z<0.6$ and 580 sources with $0.6<z<1.0$.

\subsection {Small- and large-scale environments of X-ray AGN}

It was shown, in general, that the X-ray selected AGN reside in all kinds of environments. This is   
in agreement with previous works for AGN selected in the optical, X-ray, and radio
(cf. Gilmour et al. 2009, Constantin et al. 2008, Silverman et al. 2009, Lietzen et al. 2009, 2011, Tasse et al. 2008, 2011, Melnyk et al. 2013, 
Hwang et al. 2012, Gendre et al. 2013, Manzer \& De Robertis 2014, Karouzos et al. 2014a,b, and Song et al. 2016).

On the other hand, according to the small-scale analysis, 
the Kolmogorov--Smirnov probability that the AGN overdensity distribution is consistent with the mock distribution 
is less than $10^{-3}$ in the first redshift range and for both the bright and faint environments. The
AGN$_{soft}$ and AGN$_{hard}$ (i.e. unobscured and obscured AGN), radio, and non-radio X-ray sources also prefer the overdense bright or faint environments more 
frequently than in the mock catalogues at a high level of significance ($>3\sigma$). There is also an apparent shift to higher overdensity values at lower redshifts. 
This result is in agreement with X-ray AGN environmental studies by Melnyk et al. (2013), Silverman et al. (2009), and \cite{2014MNRAS.439..861K}. 
It also agrees with some studies for optical AGN, although there are many studies of optical AGN that do not agree. 
For example, Hwang et al. (2012) and Song et al. (2016) found that the fraction of AGN in the field is higher 
than in clusters 
(see also the references in those papers), contrary to Manzer \& De Robertis (2014) for whom the AGN fraction increases with decreasing distance to the group centroid. 
However, \cite{2015MNRAS.451.1482M} and \cite{2016arXiv160505642A} 
independently found no difference in the prevalence of optical AGN (mainly type 2) in isolated and paired galaxies. 

According to our results, although the majority of AGN are located in small-scale overdensities, their luminosity does not correlate with the environmental density. 
\cite{2014MNRAS.439..861K} obtained similar results, concluding that  
high-luminosity AGN are not preferentially triggered by mergers. They concluded that AGN likely trace the overdense environment because they inhabit the most massive galaxies 
and not because they are triggered by interactions, supporting the scenario of secular AGN evolution.
Our findings are also in general agreement with recent environmental studies of optically selected AGN. 
In particular,  \cite{2015MNRAS.447..110S} used a sample of about 250,000 galaxies and showed that the
effects of the large-scale environment and galaxy interactions are minimal on both the prevalence of optically selected  HLAGN and on their luminosity,  
supporting the scenario of a secular AGN evolution. We are motivated to test the nearest neighbour effect in the XXL-N field in the future, 
where many spectroscopic redshifts have become available.

The differences between the environment of AGN$_{soft}$ (i.e. mainly unobscured or broad-line AGN, HLAGN, type 1) and AGN$_{hard}$ 
(mainly obscured AGN, narrow-line, MLAGN, type 2\footnote{This classification is not explicit.}) are also negligible. 
This is in agreement with the classical unified scheme for AGN by
Antonucci (1993). In our previous work for the XMM-LSS field (Melnyk et al. 2013), 
we found some evidence that AGN$_{hard}$ are located in more overdense regions than AGN$_{soft}$, 
although the significance level was relatively low ($<3\sigma$) and the number of sources was three times lower than in the present study. 
It is also in agreement with Gilli et al. (2009) and Strand et al. (2008).

We have to note that the considered small-scale overdensities were computed within 400 kpc, 
while other studies have reported the difference between obscured and unobscured AGN at very 
small distances, $<$30-100 kpc (see e.g. Koulouridis et al. 2006 \& 2013, Koss et al. 2012, 
Satyapal et al. 2014). Also, we do not consider  the host galaxy types in this paper.
 
We did not find that X-ray radio-selected AGN prefer denser environments than non-radio selected ones, unlike  previous studies (Lietzen et al. 2011, 
Sabater et al. 2013, Karouzos et al. 2014a \& b, Ineson et al. 2015, Argudo-Fernandez et al. 2016, and Bradshaw et al. 2011). 
However, we found that among  the X-ray selected sources, radio detections display significantly higher hardness ratios than radio undetected sources.

We did not find evidence for any influence of the large-scale ($>$ 1 Mpc) environment on X-ray or radio properties of the sources. We again note that
our assumption here was that the X-ray sources closely trace the distribution of optical galaxies. According to Song et al. (2016), the
optical quasar density changes somewhat more slowly than the galaxy density. Nevertheless, taking into account the field galaxies, those authors showed a weak positive correlation 
between the black hole mass and the large-scale environmental density, and a negative correlation between the optical luminosity and the density. As previously noted, 
due to very non-homogeneous spectroscopic coverage of the XXL-S field galaxies, we were not able to calculate the environmental density of the field galaxies.

\subsection {Large-scale structure of the XXL-S}

We find no correlation between the small-scale and large-scale environments of the X-ray sources, implying that there does not seem to be any preferable environment for X-ray sources. 
Nevertheless, they avoid the most empty regions of galaxies (voids). This is in agreement with the fact that powerful X-ray AGN are rarely observable among isolated galaxies 
in the Local Universe (Pulatova et al. 2015). In an earlier work by \cite{2007ApJ...660L..15G} it was also shown that X-ray selected AGN at $z\sim1$ avoid underdense regions at 
the 99.89\% confidence level.

We found that X-ray AGN typically reside at relatively small distances from the centres of X-ray clusters ($<$5 Mpc), 
which may refer to the absence of the `AGN suppression' effect which 
was reported in previous works by \cite{2010ApJ...714L.181K} and \cite{2014MNRAS.437.1942E}. However, our findings are in agreement with \cite{2014A&A...567A..83K}, 
where the effect of AGN suppression in X-ray clusters was not found for the poor clusters in the XMM-LSS field.
Koulouridis et al. (2016; \cite{2016A&A...592A..11K}) argue that the total number of AGN in the vicinity of three superclusters significantly exceeds 
the field expectations. 
Although superclusters represent the most extensive concentrations in the Universe, they do not represent the densest environments. 
We expect that future XXL papers dedicated to studying AGN counts in X-ray clusters will clarify this issue.

In general, there seems to be an anticorrelation between high density and AGN activity.
We found that X-ray AGN trace the distribution of X-ray clusters at a $<25-45$ Mpc scale: the fraction of agglomerates
 located in the vicinity of X-ray clusters is $\sim$1.5-2 times higher than for randomly distributed agglomerates in both considered redshift ranges.
This is in agreement with Arnold et al. (2009) and also with optical AGN studies by Hwang et al. (2012) and Song et al. (2016), where the authors show that the AGN fractions  
in the field environment are higher than in clusters. Kocevski et al. (2009) found that Seyfert-2 galaxies 
 avoid the densest regions of superclusters and are instead located in intermediate density environments. Moreover, we found that the most populated agglomerates
 are associated with supercluster candidates (see XXL Paper XX).
  We interpret our results as showing that X-ray AGN mainly reside in supercluster  filaments/field environments.

To summarise our results: 
\begin{enumerate}
 \item The large-scale environment does not correlate with any specific AGN population studied here. 
 \item Obscured/unobscured AGN, radio, and non-radio sources typically reside in small-scale overdensities, a trend which is stronger at lower redshifts. 
 \item No correlation was found between small-scale overdensities and X-ray luminosity, nor between environmental density and 
 the type of AGN. Radio sources also prefer the same locally overdense environments as non-radio AGN.
 \item A large number of AGN concentrations with two or more members correlates with the presence of X-ray galaxy clusters within $<$25-45 Mpc. 
\end{enumerate}

\begin{acknowledgements}

XXL is an international project based around an XMM Very Large Programme surveying two 25 deg$^{2}$ extragalactic fields at a depth of 
$\sim$5 $\cdot$ 10$^{-15}$ erg cm$^{-2}$ s$^{-1}$ in the [0.5-2] keV band for point-like sources. 
The XXL website is http://irfu.cea.fr/xxl. Multiband information and spectroscopic follow-up of the X-ray sources were obtained through a number of survey programmes, 
http://xxlmultiwave.pbworks.com/.

The research leading to these results has received funding from the European Union Seventh Framework Programme (FP7 2007-2013) under grant agreement 
291823 Marie Curie FP7-PEOPLE-2011-COFUND (The new International Fellowship Mobility Programme for Experienced Researchers in Croatia - NEWFELPRO). 
This article has been written as part of the project ``AGN environs in XXL'' 
which has received funding through NEWFELPRO project under grant agreement No83. 
The Saclay group acknowledges long-term support from the Centre National d'Etudes Spatiales (CNES).
EK thanks CNES and CNRS for support of post-doctoral research. VS, JD, and ID acknowledge funding by the European Unions Seventh Framework programme under
grant agreement 337595 (ERC Starting Grant, CoSMass). This research has made use of the XMM-LSS database, operated at CeSAM/LAM, Marseille, France.

\end{acknowledgements}

\end{document}